\documentclass[published]{JHEP3}

\JHEP{00(2007)000}

\JHEPspecialurl{http://jhep.sissa.it/JOURNAL/JHEP3.tar.gz}

\usepackage{epsfig,multicol,bbm}
\usepackage{amsmath,amssymb}
\usepackage{graphicx}

\newcommand\fverb{\setbox\fverbbox=\hbox\bgroup\verb}
\newcommand\fverbdo{\egroup\medskip\noindent%
			\fbox{\unhbox\fverbbox}\ }
\newcommand\fverbit{\egroup\item[\fbox{\unhbox\fverbbox}]}
\newbox\fverbbox

\title{Asymptotic freedom in inflationary cosmology with
a non-minimally coupled Higgs field}

\author{Andrei O. Barvinsky\\Theory Department, Lebedev
Physics Institute,
Leninsky Prospect 53, Moscow 119991, Russia\\

E-mail: \email{barvin@td.lpi.ru}}

\author{Alexander Yu. Kamenshchik \\Dipartimento di Fisica and INFN,
via Irnerio 46, 40126 Bologna, Italy\\
L. D. Landau Institute for
Theoretical Physics, Moscow 119334, Russia\\

E-mail: \email{kamenshchik@bo.infn.it}}

\author{Claus Kiefer\\Institut f\"ur Theoretische Physik,
Universit\"at zu K\"oln, Z\"ulpicher Strasse 77,
50937 K\"oln, Germany\\

E-mail: \email{kiefer@thp.uni-koeln.de}}

\author{Alexei A. Starobinsky\\
L. D. Landau Institute for
Theoretical Physics, Moscow 119334, Russia
\\
RESCEU, Graduate School of Science,
The University of Tokyo, Tokyo 113-0033, Japan\\

E-mail: \email{alstar@landau.ac.ru}}

\author{Christian Steinwachs\\Institut f\"ur Theoretische Physik,
Universit\"at zu K\"oln, Z\"ulpicher Strasse 77,
50937 K\"oln, Germany\\

E-mail: \email{cst@thp.uni-koeln.de}}

\received{\today} 		
\accepted{\today}		

\abstract{
We consider the renormalization group improvement in the theory of
the Standard Model Higgs boson playing the role of an inflaton with
a strong non-minimal coupling to gravity. It suggests the range of
the Higgs mass $135.6\; {\rm GeV} \lesssim M_H\lesssim 184.5\;{\rm
GeV}$ compatible with the current CMB data (the lower WMAP bound on
$n_s$), which is close to the widely accepted range dictated by the
electroweak vacuum stability and perturbation theory bounds. We find
the phenomenon of asymptotic freedom induced by this non-minimal
curvature coupling, which brings the theory to the weak coupling
domain everywhere except at the lower and upper boundary of this range.
The renormalization group running of the basic quantity
$\mbox{\boldmath $A_I$}$ -- the anomalous scaling in the
non-minimally coupled Standard Model, which analytically determines
all characteristics of the CMB spectrum -- brings $\mbox{\boldmath
$A_I$}$ to small negative values at the inflation scale. This
property is crucial for the above results and may also underlie
the formation of initial conditions for the inflationary dynamics in
quantum cosmology.
}


\begin{document} 

\section{Introduction}

The task of constructing a fundamental particle model accounting for
an inflationary scenario and its observationally consistent
predictions for the CMB
was undertaken in \cite{we-scale} and, more recently, in the series of
papers \cite{BezShap,we,BezShap1,Wil,BezShap3}. While the
relatively old work \cite{we-scale} had suggested that due to
quantum effects inflation depends not only on the
inflaton-graviton sector of the system but rather is strongly
effected by its GUT contents, the paper by Bezrukov and
Shaposhnikov \cite{BezShap} transcended this idea from the remote
land of Grand Unification Theory to the seemingly firm ground of
the Standard Model (SM) with the Higgs field playing the role of
an inflaton. This has provoked new interest in a once rather popular
\cite{Spokoiny,FM89,SBB,Unruh,we-scale,BK,KomatsuFutamase} but
then nearly forgotten model with the Lagrangian of the
graviton-inflaton sector given by
    \begin{eqnarray}
    &&{\mbox{\boldmath $L$}}(g_{\mu\nu},\varPhi)=
    \frac12\left(M_P^2+\xi|\varPhi|^2\right)R
    -\frac{1}{2}|\nabla\varPhi|^{2}
    -V(|\Phi|),  \label{inf-grav}\\
    &&V(|\Phi|)=\frac{\lambda}{4}(|\varPhi|^2-v^2)^2,\,\,\,\,
    |\varPhi|^2=\varPhi^\dag\varPhi,
    \end{eqnarray}
where $\varPhi$ is one of the scalar multiplets of the GUT-type
sector of the system, whose expectation value plays the role of an
inflaton and which has a strong non-minimal curvature coupling with
$\xi\gg 1$. Here, $M_P=m_P/\sqrt{8\pi}\approx 2.4\times 10^{18}$ GeV is a
reduced Planck mass, $\lambda$ is a quartic self-coupling of
$\varPhi$, and $v$ is a symmetry breaking scale.

The motivation for this model was  based in the early papers
\cite{Spokoiny,FM89,SBB,Unruh} on the observation that the problem
of an exceedingly small quartic coupling $\lambda\sim 10^{-13}$,
dictated by the amplitude of primordial scalar (density)
perturbations \cite{H82,S82,GP82}, can be solved by using a
non-minimally coupled inflaton with a large value of $\xi$,
because the CMB anisotropy $\Delta T/T\sim 10^{-5}$ is
proportional to the ratio $\sqrt\lambda/\xi$, rather than to
$\sqrt\lambda$ as in the minimal case. Therefore, a small $\Delta
T/T$ can be obtained even for $\lambda$ close to unity (but still
small enough to justify perturbative expansion in $\lambda$) if
$\xi\sim 10^4$.

Later the model (\ref{inf-grav}) with the GUT-type sector of matter
fields was used to generate initial conditions for inflation
\cite{we-scale} within the concept of the no-boundary \cite{noboundary}
and tunneling cosmological states \cite{L84,R84,ZS84,tunnel}. The quantum
evolution with these initial data was considered in \cite{BK,efeqmy}.
In particular, it was shown that quantum effects are critically
important for this scenario.

A similar model but with the Standard Model (SM) Higgs boson
$\varPhi$ playing the role of an inflaton instead of the abstract
GUT setup of \cite{we-scale,BK} was suggested in \cite{BezShap}.
This work advocated the consistency of the corresponding CMB data
with WMAP observations. However, it disregarded quantum
effects\footnote{This was caused by the use of a special
field-dependent renormalization scheme \cite{ShapRenorm} suppressing
the contributions of running logarithms in the effective potential.}
which were later taken into account in \cite{we} at the level of the
one-loop approximation. Application of the general method of
\cite{we-scale,BK} to the Standard Model Higgs inflation has led to
the lower bound on the Higgs mass $M_H\gtrsim 230$ GeV, originating
from the observational restrictions on the CMB spectral index
\cite{we}. However, this pessimistic conclusion, which
contradicts the widely accepted range 115 GeV$\leq M_H\leq 180$
GeV, did not take into account $O(1)$ contributions due to running
logarithms in the quantum effective action, which can
qualitatively improve the situation\footnote{The possibility of
such contributions was, in fact, mentioned in the paper \cite{we},
but the sign of their effect was not predictable at the moment of
publication.}. This was nearly simultaneously observed in
\cite{GB08,BezShap1,Wil} where the renormalization group (RG)
improvement of the one-loop results of \cite{we} was considered
and it was found that the Higgs-mass range compatible with the CMB
data almost coincides with the conventional one.

In contrast to the numerical and, therefore, not entirely
transparent analysis of \cite{BezShap1,Wil}, we develop here a
method which retains the analytical formalism of \cite{we} by
incorporating the technique of running coupling constants. This
supports the results of \cite{we} and, moreover, explicitly
reveals a mechanism which brings the CMB compatible range of the
Higgs mass closer to the conventional Standard Model domain. This
mechanism also explains the perfect efficiency of the perturbation
theory at energy scales of the inflationary stage; since this is
analogous to asymptotic freedom in QCD, we call it here asymptotic
freedom, too. This mechanism is mediated by the
effect of longitudinal (virtual) gravitons; gravitation
thus serves as a regulator making the Standard Model sector softer.

The organization of the paper is as follows. In Sects. 2 and 3 we
discuss the notion of anomalous scaling and recapitulate the results
of the one-loop approximation of \cite{we}. In Sect. 4 we develop
the RG improvement for the effective action of the theory, present
beta functions for running couplings with a special emphasis on loop
contributions of Higgs particles and Goldstone modes in the presence
of the non-minimal coupling to graviton \cite{BK,we,Wil}. In Sect. 5
we prove the applicability of the analytic method of \cite{we} for
the CMB dynamics at the inflationary stage, which is based on the shortness
of this stage compared to the post-inflationary evolution to the
present-day electroweak vacuum. Sect. 6 contains results of
numerical simulations for the RG flow of coupling constants, the
anomalous scaling ${\mbox{\boldmath $A$}}(t)$ in particular, which
impose CMB and Standard Model bounds on the Higgs mass and reveal the
asymptotic-freedom mechanism based on the behavior of
${\mbox{\boldmath $A$}}(t)$ and $\lambda(t)$. In concluding Sect. 7,
we discuss the limitations of the obtained results associated with
gauge and parametrization dependence of the underlying formalism and
suggest the method of their resolution. We also compare in the context
of these issues our conclusions with the results of \cite{BezShap3}
and \cite{Wil}. We show that the RG treatment of the model is
essentially more sensitive to the CMB bounds than in
\cite{BezShap3}, especially at the upper limit of the Higgs mass
range $\simeq 185$ GeV. We also find that the RG improvement makes
the scenario realistic in which the initial conditions for
inflation are generated from quantum cosmology in the form of a sharp
probability 
peak of the tunneling wavefunction \cite{L84,R84,ZS84,tunnel}.
The Appendix contains the derivation of SM beta functions in the
presence of a strong non-minimal curvature coupling of the Higgs
field.

\section{Anomalous scaling parameter in theories with a large
non-minmal curvature coupling} The usual understanding of
non-renormalizable theories is that renormalization of
higher-dimensional operators does not effect the renormalizable
sector of low-dimensional operators, because the former ones are
suppressed by powers of a cutoff -- the Planck mass $M_P$
\cite{Weinberg}. Therefore, beta functions of the Standard Model
sector are not expected to be modified by graviton loops. The
situation with the non-minimal coupling is more subtle. Due to the
mixing of the Higgs scalar field with the longitudinal part of
gravity in the kinetic term of the Lagrangian (\ref{inf-grav}), an
obvious suppression of pure graviton loops by the effective Planck
mass, $M_P^2+\xi\varphi^2\gg M_P^2$, for large $\xi$ proliferates
to the sector of the Higgs field, so that certain parts of beta
functions, which induce Landau poles, are strongly damped by large
$\xi$ \cite{Wil}. Therefore, running coupling constants like
$\lambda(t)$ turn out to be small or remain finite at the
inflation scale. In particular, a special combination of coupling
constants $\mbox{\boldmath$A$}$ which we call {\em anomalous
scaling} \cite{we-scale} becomes very small and reduces the
CMB-compatible Higgs-mass bound. The importance of $\mbox{\boldmath$A$}$
follows from the fact observed in \cite{we-scale,BK,we} that due
to large $\xi$, quantum effects and their CMB manifestation are
universally determined by $\mbox{\boldmath$A$}$. The nature of
this quantity, which was initially introduced in the context of a
generic gauge theory non-minimally coupled to gravity
\cite{we-scale}, is as follows.

Let the model contain in addition to (\ref{inf-grav}) also a set
of scalar fields $\chi$, vector gauge bosons $A_\mu$ and spinors
$\psi$, which have a typical interaction with $\varPhi$ dictated by
the local gauge invariance. If we denote by $\varphi$ the inflaton
-- the only nonzero component of the mean value of $\varPhi$ in the
cosmological state, $\varphi^2=\varPhi^\dag\varPhi$, then the
non-derivative part of the interaction of these particles with
$\varphi$ can be schematically written down as
    \begin{eqnarray}
    {\mbox{\boldmath $L$}}_{\rm int}
    =-\sum_{\chi}\frac12 \lambda_{\chi}
    \chi^2\varphi^2
    -\sum_{A}\frac12 g_{A}^2A_{\mu}^2\varphi^2-
        \sum_{\psi}y_{\psi}\varphi\bar\psi\psi,    \label{interaction}
    \end{eqnarray}
where $\lambda_\chi$, $g_A$ and $y_\psi$ are the relevant constants
of quartic, gauge, and Yukawa couplings.

On the background of a slowly varying inflaton $\varphi$,
these terms generate masses $m(\varphi)$ for all these particles,
which are proportional to $\varphi$, $m(\varphi)\sim\varphi$, and
which in turn generate at one-loop order the Coleman-Weinberg
potential
    \begin{eqnarray}
    &&\sum_{
    \rm particles}
    (\pm 1)\,\frac{m^4(\varphi)}{64\pi^2}
    \,\ln\frac{m^2(\varphi)}{\mu^2}
    =\frac{\lambda\mbox{\boldmath$A$}}{128\pi^2}
    \,\varphi^4
    \ln\frac{\varphi^2}{\mu^2}+...\; .  \label{Aviamasses}
    \end{eqnarray}
Here, the summation over particles includes the statistics as well as
the sum over polarizations, and the overall coefficient $\mbox{\boldmath$A$}$
schematically reads
    \begin{eqnarray}
    {\mbox{\boldmath $A$}} = \frac{2}{\lambda}
    \left(\sum_{\chi} \lambda_{\chi}^{2}
    + 3 \sum_{A} g_{A}^{4} - 4
    \sum_{\psi} y_{\psi}^{4}\right),   \label{A}
    \end{eqnarray}
where the coefficients 1, 3 and 4 in front of the sums are the
number of degrees of 
freedom of a real scalar field, a massive vector field, and a charged
Dirac spinor\footnote{For a generic model,
$(\lambda_\chi,g_A^2,y_\psi)$ comprise coupling constant matrices
which give rise to mass matrices and their traces in the actual
expression for $\mbox{\boldmath $A$}$.}. This quantity is, of
course, well known, because as a coefficient of the logarithm it
comprises the ultraviolet renormalization of $\lambda$ and
constitutes the conformal anomaly, or anomalous scaling,
associated with the normalization scale $\mu$ in
(\ref{Aviamasses}). It includes the one-loop contributions of all
particles except the graviton and the inflaton field $\varphi$
itself, because their contributions are strongly suppressed by
inverse powers of $\xi\gg 1$ \cite{BK,we,Wil} -- a property
which will be discussed in much detail below.

The role of (\ref{A}) is two-fold. First, in the context of the
no-boundary and tunneling initial conditions 
\cite{noboundary,L84,R84,ZS84,tunnel} extended to the one-loop level
\cite{we-scale}, it appears in the distribution function of the
quasi-de~Sitter cosmological instantons $\rho(\varphi)$ describing a
quantum distribution of cosmological models with different initial
$\varphi$,
        \begin{eqnarray}
        \rho(\varphi)\sim
        \left(\frac\varphi\mu\right)
        ^{\textstyle-6\mbox{\boldmath$A$}\xi^2/\lambda}. \label{rho}
        \end{eqnarray}
This scaling (which gave rise to the name ``anomalous scaling'') makes this
distribution normalizable at $\varphi\to\infty$ for positive
$\mbox{\boldmath$A$}$ \cite{norm} and, moreover, for the case of the
tunneling cosmological wavefunction it generates a sharp probability
peak in $\rho(\varphi)$ at the following value of the inflaton field:
the quantum scale of inflation derived in \cite{we-scale},
$\varphi_I^2=64\pi^2 M_P^2/\xi\mbox{\boldmath$A$}$.

Secondly, the anomalous scaling for $\xi\gg 1$ determines
the quantum rolling force in the effective equation of the
inflationary dynamics \cite{BK,efeqmy} and, consequently, yields the
parameters of the CMB generated during inflation \cite{we}. These
parameters explicitly depend on $\mbox{\boldmath$A$}$ and the
e-folding number $N$ of the first horizon crossing by the primordial
cosmological perturbation of a given wavelength. This dependence for
the spectral index $n_s$, which belongs to the interval $0.94
<n_s(k_0)<0.99$ (the combined WMAP+BAO+SN data at the pivot point
$k_0=0.002$ Mpc$^{-1}$ corresponding to $N\simeq 60$
\cite{WMAPnorm,WMAP}), gives the range of the anomalous
scaling $-12< \mbox{\boldmath$A$}<14$ \cite{we}.

On the other hand, in the Standard Model $\mbox{\boldmath$A$}$ is
expressed in terms of the masses of the heaviest particles -- $W^\pm$
boson, $Z$ boson and top quark,
    \begin{eqnarray}
    &&m_W^2=\frac14\,g^2\,\varphi^2,\;\;
    m_Z^2=\frac14\,(g^2+g'^2)\,\varphi^2,\;\;
    m_t^2=\frac12\,y_t^2\,\varphi^2,               \label{masses}
    \end{eqnarray}
and the mass of three Goldstone modes
    \begin{eqnarray}
    m_G^2=\frac{V'(\varphi)}\varphi=\lambda(\varphi^2-v^2)
    \simeq \lambda\varphi^2.               \label{Goldmass}
    \end{eqnarray}
Here, $g$ and $g'$ are the $SU(2)\times U(1)$ gauge couplings, $g_s$
is the $SU(3)$ strong coupling and $y_t$ is the Yukawa coupling for
the top quark. We work in the transversal Landau gauge with no
contribution to $m_G^2$ from the gauge-fixing term. At the inflation
stage with $\phi^2\gg v^2$, the Goldstone mass $m_G^2$ is
non-vanishing in contrast to its zero on-shell value in the electroweak vacuum
\cite{WeinbergQFT}. It is again important to emphasize that the
Higgs particle itself with the mass
    \begin{eqnarray}
    m_H^2=V''(\varphi)=\lambda(3\varphi^2-v^2)
    \simeq 3\lambda\varphi^2               \label{Higgsmass}
    \end{eqnarray}
does not contribute to ${\mbox{\boldmath $A$}}$, because its
contribution is suppressed at the inflation
scale by a small factor $\sim1/\xi^2$; this is shown in the Appendix.

Equation (\ref{Aviamasses}) then gives the expression
   \begin{equation}
    {\mbox{\boldmath $A$}} =
    \frac3{8\lambda}\Big(2g^4 +
    \big(g^2 + g'^2\big)^2- 16y_t^4 \Big)+6\lambda.   \label{A0}
    \end{equation}
In the conventional range of the Higgs mass 115 GeV$\leq M_H\leq$
180 GeV \cite{particle}, this quantity is at the electroweak scale
in the range $-48<\mbox{\boldmath$A$}<-20$, which strongly
contradicts the CMB range given above and violates as well the
normalizability of the distribution function (\ref{rho}). These two
ranges can be brought together only by the price of raising
$\lambda\simeq M_H^2/2v^2$, that is, by increasing the Higgs mass to
230 GeV \cite{we}.

Salvation comes, however, from the observation that the RG running of coupling
constants is strong enough and drives ${\mbox{\boldmath $A$}}$  to
a range compatible with the CMB data {\em and} the conventional range
of the Higgs mass. Thus, the formalism
of \cite{we} stays applicable but with the electroweak
${\mbox{\boldmath $A$}}$ replaced by the running ${\mbox{\boldmath
$A$}}(t)$,
    \begin{equation}
    {\mbox{\boldmath $A$}}(t) =
    \frac3{8\lambda(t)}\Big(2g^4(t)
    +\big(g^2(t)+g'^2(t)\big)^2
    -16y_t^4(t)\Big)+6\lambda(t).   \label{A-run}
    \end{equation}
Here, $t=\ln(\varphi/\mu)$ is the running scale of the
RG improvement of the effective potential
\cite{ColemanWeinberg}, where $\mu$ is a normalization point which
we choose to coincide with the top quark mass $\mu=M_t$ (we denote
physical (pole) masses by capital letters in contrast to the running
masses (\ref{masses}) above). In the following we shall present the
details of this mechanism.

\section{One-loop approximation}
The inflationary model with a non-minimally coupled Higgs-inflaton
was considered in the one-loop approximation in \cite{we}. The
low-derivative part of its effective action (appropriate for the
inflationary slow-roll scenario),
    \begin{equation}
    S[g_{\mu\nu},\varphi]=\int d^{4}x\,g^{1/2}
    \left(-V(\varphi)+U(\varphi)\,R(g_{\mu\nu})-
    \frac12\,G(\varphi)\,(\nabla\varphi)^2\right)\ ,   \label{effaction}
    \end{equation}
contains coefficient functions which in the one-loop
approximation read as
    \begin{eqnarray}
    &&V(\varphi)=\frac\lambda{4}(\varphi^2-\nu^2)^2+
    \frac{\lambda\varphi^4}{128\pi^2}
    \mbox{\boldmath$A$}
    \ln\frac{\varphi^2}{\mu^2},             \label{effpot}\\
    &&U(\varphi)=
    \frac12(M_P^2+\xi\varphi^{2})+
    \frac{\varphi^2}{32\pi^2}\left(\mbox{\boldmath$C$}
    \ln\frac{\varphi^2}{\mu^2}+D
    \right),                               \label{effPlanck}\\
    &&G(\varphi)=1+\frac{1}{32\pi^2}\left(F
    \ln\frac{\varphi^2}{\mu^2}
    +E\right).                              \label{phirenorm}
    \end{eqnarray}
These functions contain numerical coefficients
$\mbox{\boldmath$A$},\mbox{\boldmath$C$},D,F,E$ determined by
contributions of quantum loops of all particles and include the
dependence on the UV normalization scale $\mu$. For $\xi\gg 1$ the
inflationary stage and the corresponding CMB parameters critically
depend only on the anomalous scaling $\mbox{\boldmath$A$}$ defined
by (\ref{Aviamasses}) \cite{we-scale,BK,efeqmy,we} and the part of
$\mbox{\boldmath$C$}$ which is linear in $\xi$.
As shown in Appendix A, the
latter is contributed by the Goldstone mass $m_G^2$ and equals
        \begin{eqnarray}
        \mbox{\boldmath$C$}=3\xi\lambda+O(\xi^0). \label{C}
        \end{eqnarray}
In fact, this expression yields a quantum correction to the
non-minimal coupling $\xi$, as can be seen from (\ref{effPlanck}).
 Other coefficients, the normalization
scale $\mu$ inclusive, are irrelevant in the leading order of the
slow roll expansion.

Inflation and its CMB are easy to analyze in the Einstein frame of
fields $\hat g_{\mu\nu}$, $\hat\varphi$, which is related to the Jordan frame
used in (\ref{effaction}) by the equations
        \begin{eqnarray}
        \hat g_{\mu\nu}=\frac{2U(\varphi)}
        {M_P^2}g_{\mu\nu},\,\,\,\,
        \left(\frac{d\hat\varphi}{d\varphi}\right)^2
        =\frac{M_P^2}{2}\frac{GU+3U'^2}{U^2}.      \label{Eframe}
        \end{eqnarray}
The action (\ref{effaction}) in the Einstein frame, $\hat S[\hat
g_{\mu\nu},\hat\varphi]=S[g_{\mu\nu},\varphi]$, has a minimal
coupling, $\hat U=M_P^2/2$, a canonically normalized inflaton field,
$\hat G=1$, and a new inflaton potential,
        \begin{eqnarray}
        \hat{V}(\hat\varphi)=\left.\left(\frac{M_P^2}{2}\right)^2
        \frac{V(\varphi)}{U^2(\varphi)}
        \,\right|_{\,\varphi=\varphi(\hat\varphi)}~.     \label{hatV}
        \end{eqnarray}

In view of (\ref{effpot}), (\ref{effPlanck}) and (\ref{C}), at the
inflation scale with $\varphi>M_P/\sqrt{\xi}\gg v$ and $\xi\gg 1$,
this potential reads
        \begin{eqnarray}
        \hat{V}=\frac{\lambda
        M_P^4}{4\,\xi^2}\,\left(1-\frac{2M_P^2}{\xi\varphi^2}+
        \frac{\mbox{\boldmath$A_I$}}{16\pi^2}
        \ln\frac{\varphi}{\mu}\right),            \label{hatVbigphi}
        \end{eqnarray}
where the parameter $\mbox{\boldmath$A_I$}$ represents the anomalous
scaling (\ref{A0}) modified by quantum corrections to $U$ -- the
renormalization of the non-minimal term (\ref{C}),
       \begin{eqnarray}
        \mbox{\boldmath$A_I$}=\mbox{\boldmath$A$}-12\lambda=
        \frac3{8\lambda}\Big(2g^4 +
        \big(g^2 + g'^2\big)^2- 16y_t^4 \Big)-6\lambda.  \label{AI}
        \end{eqnarray}
This quantity -- which we will call {\em inflationary anomalous
scaling} -- enters the expressions for the standard slow-roll parameters,
    \begin{eqnarray}
    &&\hat\varepsilon \equiv\frac{M_P^2}2\left(\frac1{\hat
    V}\frac{d\hat V}{d\hat\varphi}\right)^2=
    \frac43\left(
    \frac{M_P^2}{\xi\,\varphi^2}+
    \frac{\mbox{\boldmath$A_I$}}{64\pi^2}\!\right)^2,  \label{varepsilon}\\
    &&\hat\eta\equiv \frac{M_P^2}{\hat V}
    \frac{d^2\hat V}{d\hat\varphi^2}
    =-\frac{4M_P^2}{3\xi\varphi^2}~,      \label{eta}
    \end{eqnarray}
and ultimately determines all the characteristics of inflation. In
particular, the smallness of $\hat\varepsilon$ yields the range of the
inflationary stage $\varphi>\varphi_{\rm end}$, terminating at a
value of $\hat\varepsilon$ which for elegance of the formalism we
chose to be $\hat\varepsilon_{\rm end}=3/4$. Then the inflaton value
at the exit from inflation equals $\varphi_{\rm end}\simeq
2M_P/\sqrt{3\xi}$ under the natural assumption that the perturbation
loop expansion is applicable for $\mbox{\boldmath$A_I$}/64\pi^2\ll
1$. The duration of inflation which starts at $\varphi$ takes in
units of the scale factor e-folding number $N$ a particularly simple
form \cite{we},
    \begin{eqnarray}
    &&\frac{\varphi^2}{\varphi_I^2}
    =e^x-1,                           \label{xversusvarphi}\\
    &&\varphi_I^2=\frac{64\pi^2 M_P^2}{\xi
    \mbox{\boldmath$A_I$}},               \label{quantumscale}
    \end{eqnarray}
where the quantum scale of inflation $\varphi_I$ and a special
parameter $x$ containing the e-folding number,
    \begin{eqnarray}
    x\equiv\frac{N
    \mbox{\boldmath$A_I$}}{48\pi^2},           \label{x}
    \end{eqnarray}
directly involve the anomalous scaling $\mbox{\boldmath$A_I$}$ which
is essentially a quantum quantity.

This relation determines the Fourier power spectrum for the scalar
metric perturbation $\zeta$,
$\Delta_{\zeta}^2(k) \equiv <k^3\zeta_{{\bf k}}^2>
= \hat V/24\pi^2M_P^4\hat \varepsilon$,
where the right-hand side is taken
at the  first horizon crossing, $k=aH$, relating the comoving
perturbation wavelength $k^{-1}$ to the e-folding number $N$,
    \begin{eqnarray}
    \Delta_{\zeta}^2=
    \frac{N^2}{72\pi^2}\,\frac\lambda{\xi^2}\,
    \left(\frac{e^x-1}{x\,e^x}\right)^2.       \label{zeta}
    \end{eqnarray}
The CMB spectral index $n_s\equiv 1+d\ln\Delta_{\zeta}^2/d\ln
k=1-6\hat\varepsilon+2\hat\eta$, the tensor to scalar ratio
$r=16\hat\varepsilon$ and the spectral index running $\alpha\equiv
dn_s/d\ln k\simeq -dn_s/dN$ correspondingly read as
    \begin{eqnarray}
    &&n_s=
    1-\frac{2}{N}\, \frac{x}{e^x-1}~,           \label{ns}\\
    &&r=\frac{12}{N^2}\,
    \left(\frac{x e^x}{e^x-1}\right)^2~,          \label{r}\\
    &&\alpha=
    -\frac{2}{N^2}\,\frac{x^2e^x}{(e^x-1)^2}.     \label{running}
    \end{eqnarray}
Note that for $|x|\ll 1$ these predictions exactly coincide with
those \cite{MC81,S83} of the $f(R)=M_P^2(R+R^2/6M^2)/2$ inflationary
model \cite{S80} with the scalar particle (scalaron) mass
$M=M_P\sqrt \lambda/\sqrt 3 \xi$.

With the spectral index constraint $0.94 <n_s(k_0)<0.99$ at the
pivot point $k_0=0.002$ Mpc$^{-1}$ corresponding to $N\simeq 60$
\cite{WMAPnorm,WMAP} these relations immediately give the range of
anomalous scaling $-12< \mbox{\boldmath$A_I$}<14$. As mentioned
above, this contradicts the Standard Model bound
$-48<\mbox{\boldmath$A_I$}<-20$ existing in the conventional range
of the Higgs mass 115 GeV$\leq M_H\leq$ 180 GeV
\cite{particle}.\footnote{These estimates were obtained in \cite{we}
for the parameter $\mbox{\boldmath$A$}$ disregarding the
contribution of Goldstone modes ($+6\lambda$ and $-6\lambda$ terms
in (\ref{A0}) and (\ref{AI}), respectively), but numerically with
$\lambda\lesssim 1$ the inclusion of this contribution leads to
qualitatively the same conclusions.}

The limitation of this approximation follows from the fact that,
despite the smallness of $\mbox{\boldmath$A_I$}/64\pi^2$, the
running of the logarithm from the electroweak scale $\varphi=v$
(where the Higgs and other particle masses are determined) to the
inflation scale $\varphi\simeq|\varphi_I|$ is big enough and changes
the predictions by $O(1)$ factors. Indeed, with the values of
$\xi\sim 10^4$ and $|\mbox{\boldmath$A_I$}|\sim 10$ dictated by the
results of \cite{we}, the estimate for the logarithmic running
$\ln(|\varphi^2_I|/v^2)\sim 60$ yields the contribution
    \begin{eqnarray}
    \frac{\mbox{\boldmath$A_I$}}{64\pi^2}
    \ln\frac{|\varphi^2_I|}{v^2}\sim 2=O(1),
    \end{eqnarray}
which requires resummation within the RG improvement. The necessity
of this improvement was mentioned in \cite{we}, and it was realized
in \cite{BezShap1,Wil}. Below we develop a combined numerical and
analytical scheme of this improvement, which recovers the above
formalism of Eqs. (\ref{xversusvarphi})--(\ref{running})
incorporating running coupling constants.

\section{RG improvement}
According to the Coleman--Weinberg technique \cite{ColemanWeinberg},
the one-loop RG improved effective action has the form
(\ref{effaction}) with the coefficients
    \begin{eqnarray}
    &&V(\varphi)=
    \frac{\lambda(t)}{4}\,Z^4(t)\,\varphi^4,  \label{RGeffpot}\\
    &&U(\varphi)=
    \frac12\Big(M_P^2
    +\xi(t)\,Z^2(t)\,\varphi^{2}\Big),      \label{RGeffPlanck}\\
    &&G(\varphi)=Z^2(t).            \label{phirenorm1}
    \end{eqnarray}
Here, $t=\ln(\varphi/M_t)$ is the running RG scale\footnote{Application
of the Coleman--Weinberg technique removes the ambiguity in the
choice of the RG scale in cosmology -- an issue discussed in
\cite{Woodard}.}, and the running couplings $\lambda(t)$, $\xi(t)$
and the field renormalization $Z(t)$ incorporate a summation of powers
of logarithms and belong to the solution of the RG equations
    \begin{eqnarray}
    &&\frac{d g_i}{d t}
    =\beta_{g_i},\,\,\,\,
    g_i=(\lambda,\xi,g,g',g_s,y_t),           \label{renorm0}\\
    &&\frac{dZ}{d t}
    =\gamma Z                                \label{renormZ}
    \end{eqnarray}
for the full set of coupling constants in the ``heavy'' sector of the
model with the corresponding beta functions $\beta_{g_i}$ and the
anomalous dimension $\gamma$ of the Higgs field.\footnote{Note that
running couplings in the effective potential theory are the
characteristics of the RG equation in partial derivatives for this
potential, rather than the usual running coupling constants directly
run by $\beta$'s. Thus the right-hand sides of RG
eqs.(\ref{renorm0})-(\ref{renormZ}) should contain the factor
$1/(1-\gamma)$ \cite{ColemanWeinberg} which we discard in the
one-loop RG improvement.} This sector of the Standard Model includes the
$SU(2)\times U(1)$ gauge couplings $g$ and $g'$, the $SU(3)$ strong
coupling $g_s$, and the Yukawa coupling $y_t$ for the top quark.

An important subtlety with these $\beta$ functions is the effect of
the non-minimal curvature coupling of the Higgs field. For large $\xi$
the kinetic term of the tree-level action has a strong mixing
between the graviton $h_{\mu\nu}$ and the quantum part of the Higgs
field $\sigma$ on the background $\varphi$. Symbolically it has the
structure
\[ (M_P^2+\xi^2\varphi^2)h\nabla\nabla
h+\xi\varphi\sigma\nabla\nabla h+\sigma\Box\sigma,\] which yields a
propagator whose elements are suppressed by a small $1/\xi$-factor
in all blocks of the $2\times2$ graviton-Higgs sector. For large
$\varphi\gg M_P/\sqrt\xi$, the suppression of pure graviton loops
is, of course, obvious because the effective Planck-mass squared
exceeds by far the standard Planck-mass squared, $M_P^2+\xi\varphi^2\gg
M_P^2$. Due to the mixing, this suppression proliferates to the full
graviton-Higgs sector of the theory. To make this statement
quantitative, one can go over to the Einstein frame (\ref{Eframe}) in
which the propagator is diagonal and canonically normalized in the
space of perturbations of $\hat g_{\mu\nu}$ and $\hat\varphi$ (in
the background covariant DeWitt gauge). The above propagator can be
uplifted back to the Einstein frame by matrix multiplication with
the Jacobian matrices $\partial(g_{\mu\nu},\varphi)/\partial(\hat
g_{\alpha\beta},\hat\varphi)$. Its Higgs-Higgs block, in particular,
gets suppressed by the factor $s(\varphi)$
    \begin{eqnarray}
    &&\left(\frac{\partial\varphi}{\partial\hat\varphi}\right)^2
    \frac1{\hat g^{1/2}(\hat\Box-\hat m_H^2)}
    =
    \frac{s(\varphi)}{g^{1/2}(\Box-m_H^2)},                \label{mod}\\
    &&s(\varphi)\equiv
    \frac{U}{GU+3U'^2}=
    \frac{M_P^2+\xi\varphi^2}
    {M_P^2+(6\xi+1)\xi\varphi^2},         \label{s}
    \end{eqnarray}
where we took into account the scaling $\hat g^{1/2}(\hat\Box-\hat
m_H^2)=(2U/M_P^2)g^{1/2}(\Box-m_H^2)$ under the conformal
transformation (\ref{Eframe}) (disregarding spacetime gradients of
$\varphi$ and bearing in mind conformal rescaling of particle masses
\cite{we}). The origin of this suppression factor $s(\varphi)$
within the original Jordan frame is clearly shown in Appendix A.

This mechanism was first understood in \cite{SBB} when estimating
the CMB generation in a non-minimal inflationary model. It was
considered in \cite{our-ren} in the context of the generalized RG
approach to nonrenormalizable theories. Also it justifies the
omission of graviton loops \cite{BK,efeqmy} and modifies the beta
functions of the SM sector of this theory \cite{Wil} at the high
energy scale relevant for inflation. This modification is due to
the suppression of every Higgs propagator (\ref{mod}) by the
factor $s(\varphi)$ \cite{Wil}, which is very small for
$\varphi\gg M_P/\sqrt\xi$, $s\simeq 1/6\xi$, but tends to one in
the vicinity of the electroweak scale $v\ll M_P/\xi$. Such a
modification justifies, in fact, the extension beyond the scale
$M_P/\xi$ interpreted in \cite{BarbEsp,BurgLeeTrott} as a natural
validity cutoff of the theory\footnote{The notion of a cutoff
depends on the type of perturbation theory used. Typically it is
determined for the case when all dimensional quantities -- fields
and their derivatives -- are considered on equal footing and
treated perturbatively. However, a nonperturbative summation of
powers of the field for low-derivative processes (achieved by the
conformal transformation to the Einstein frame) allows one to go
beyond such a cutoff, and this is exactly the situation with the
scales above $M_P/\xi$, see e.g. \cite{BezShap2}.}.

Of course, the factor $s(\varphi)$ makes the counterterms of the
theory nonpolynomial in $\varphi$, and the theory as a whole becomes
nonrenormalizable. One might think that this completely invalidates
the construction of the usual RG improvement and requires the
generalization of the latter to RG with an infinite set of charges
(perhaps of the functional nature like in \cite{Kazakov,our-ren}).
However, there exists a shortcut to a simpler formulation based on
the structure of the factor (\ref{s}). In the logarithmic scale of
the variable $t=\ln(\varphi/\mu)$, strongly compressing the
transition domain $M_P/\xi\lesssim\varphi\lesssim M_P/\sqrt\xi$ for
$s(\varphi)$ from 1 to $1/6\xi\simeq 0$, this factor looks very much
like a step function $\theta(t_0-t)$,
$t_0\simeq\ln(\varphi_0/\mu)=\ln(M_P/\sqrt{6}\xi\mu)$. So it is
nearly constant for $\varphi<\varphi_0$ and $\varphi>\varphi_0$.
Therefore, in both phases of such an approximation the theory is
renormalizable but has different running of coupling
constants.\footnote{This is analogous to the approach of
\cite{BezShap3} where the model was approximated by matching the
usual low-energy SM phase with the chiral phase of the SM at
the inflation energy scale. This picture, however, takes place in
the Einstein frame of the theory and is aggravated by the problem of
transition between two different parameterizations of one quantum
theory --- the Cartesian coordinates in the space of the Higgs
multiplet in the low-energy phase versus spherical coordinates in the
chiral phase of the SM. This transition is physically nontrivial and is
discussed below in more detail.} This qualitatively justifies the
use of beta functions modified by $s$-factors smoothly interpolating
between these two phases. Below we will see that numerical results
for smooth and step function $s$-factors nearly coincide with one
another.

There is an important subtlety with the modification of beta
functions, which was disregarded in \cite{Wil} (and the first
version of this paper). The mnemonic rule of
associating the factor $s(\varphi)$ with every propagator of the
Higgs multiplet, as suggested in \cite{Wil}, turned out to be
incorrect. Goldstone modes, in contrast to the Higgs particle, are
not coupled to curvature, and they do not have a kinetic term
mixing with gravitons \cite{BezShap3}. Therefore, their
contribution is not suppressed by the $s$-factor of the above
type. Separation of Goldstone contributions from the Higgs contribution
leads to the following modification of the one-loop beta
functions, which is essentially different from that of \cite{Wil}
(cf. also \cite{Clarcketal}):
    \begin{eqnarray}
    &&\beta_{\lambda} = \frac{\lambda}{16\pi^2}
    \left(18s^2\lambda
    +{\mbox{\boldmath $A$}}(t)\right)
    -4\gamma\lambda,                           \label{beta-lambda}\\
    &&\beta_{\xi} =
    \frac{6\xi}{16\pi^2}(1+s^2)\lambda
    -2\gamma\xi,                 \label{beta-xi}\\
    &&\beta_{y_t} = \frac{y_t}{16\pi^2}
    \left(-\frac{2}{3}g'^2
    - 8g_s^2 +\left(1+\frac{s}2\right)y_t^2\right)
    -\gamma y_t,                                    \label{beta-y}
\end{eqnarray}
and 
\begin{eqnarray}
    &&\beta_{g} = -\frac{39 - s}{12}
    \frac{g^3}{16\pi^2},                     \label{beta-g}\\
    &&\beta_{g'} =
    \frac{81 + s}{12} \frac{g'^3}{16\pi^2},  \label{beta-g1}\\
    &&\beta_{g_s} =
    -\frac{7 g_s^3}{16\pi^2}.                    \label{beta-gs}
    \end{eqnarray}
Here, the anomalous dimension of the Higgs field $\gamma$ is given by
a standard expression in the Landau gauge,
    \begin{eqnarray}
    \gamma=\frac1{16\pi^2}\left(\,\frac{9g^2}4
    +\frac{3g'^2}4 -3y_t^2\right),                  \label{gamma}
    \end{eqnarray}
the anomalous scaling ${\mbox{\boldmath $A$}}(t)$ is defined by
(\ref{A-run}), and we have retained only the leading terms in $\xi\gg 1$.
It will be important in what follows that this anomalous scaling
contains the Goldstone contribution $6\lambda$, so that the full
$\beta_\lambda$ in (\ref{beta-lambda}) has a $\lambda^2$-term
unsuppressed by $s(\varphi)$ at large scale $t=\ln(\varphi/\mu)$. In
Appendix A we derive the Goldstone and Higgs contributions to
$\beta_\lambda$ and $\beta_\xi$, which demonstrates their different
suppression mechanisms under non-minimal coupling with curvature.

\section{Inflationary stage versus post-inflationary running}
As we will see, the inflationary stage in units of Higgs-field
e-foldings is very short. This allows us to make one more
approximation: we shall consider the solutions of the RG equations only up to
terms linear in $\Delta t\equiv t-t_{\rm end}=
\ln(\varphi/\varphi_{\rm end})$, where we choose as the initial data
point the end of inflation $t_{\rm end}$. This approximation will be
justified later in most of the Higgs-mass range compatible with the
CMB data.

Thus, we use beta functions (\ref{beta-lambda}) and (\ref{beta-xi})
with $s=0$ to obtain
    \begin{eqnarray}
    &&\lambda(t) = \lambda_{\rm end}\left(1
    - 4\gamma_{\rm end}\Delta t
    +\frac{\mbox{\boldmath $A$}(t_{\rm end})}{16\pi^2}\,
    \Delta t\right),                             \label{lambda-lin}\\
    &&\xi(t) = \xi_{\rm end}\Big(1
    -2\gamma_{\rm end}\Delta t
    +\frac{6\lambda}{16\pi^2}\Delta t\Big).                \label{xi-lin}
    \end{eqnarray}
Here, $\lambda_{\rm end}$, $\gamma_{\rm end}$, $\xi_{\rm end}$ are
determined at $t_{\rm end}$ and ${\mbox{\boldmath $A$}}_{\rm
end}={\mbox{\boldmath $A$}}(t_{\rm end})$ is also the particular value
of the running anomalous scaling (\ref{A-run}) at the end of
inflation.

On the other hand, the RG improvement of the effective action
(\ref{RGeffpot})--(\ref{phirenorm1}) implies that this action
coincides with the tree-level action with running couplings as
functions of $t=\ln(\varphi/\mu)$ for a new field
    \begin{eqnarray}
    \phi=Z(t)\,\varphi.     \label{phi}
    \end{eqnarray}
This is because the running of $Z(t)$ is slow,
$\partial_\mu\phi=Z(1+\gamma)\partial_\mu\varphi\simeq
Z\partial_\mu\varphi$, and the kinetic term of the effective action
in terms of $\phi$ gets canonically normalized. Then, in view of
(\ref{RGeffpot})--(\ref{RGeffPlanck}) the RG improved potential for
this field takes at the inflation stage the form
    \begin{equation}
    \hat V = \left(\frac{M_P^2}2\right)^2
    \frac{V}{U^2}
    \simeq M_P^4\frac{\lambda_{\rm end}}{4\xi^2_{\rm end}}
    \left(1-\frac{2M_P^2}{\xi_{\rm end}\phi^2} +
    \frac{{\mbox{\boldmath $A_I$}}(t_{\rm end})}{16\pi^2}
    \,\ln\frac\phi{\phi_{\rm end}}\right),      \label{eff-pot}
    \end{equation}
where for the same reason we disregarded the running of $Z$ in
$\ln(Z/Z_{\rm end})$. This is nothing but the one-loop potential
(\ref{hatVbigphi}) for the field $\phi$ with a particular choice of
the normalization point $\mu=\phi_{\rm end}$ and the couplings
replaced by the values of the running ones at $t_{\rm end}$.

This means that the formalism of \cite{we} can directly be applied
to determine the parameters of the CMB. They are mainly determined
by the anomalous scaling ${\mbox{\boldmath $A_I$}}$, but now this
quantity should be taken at $t_{\rm end}$ rather than at the
electroweak scale $t=0$. This can solve the problem of matching the
electroweak range of the anomalous scaling $-48 < {\mbox{\boldmath
$A_I$}}(0)< -20$ (corresponding to 115 GeV $\leq M_H \leq 180$ GeV)
with the range required by the CMB data, $-12.4<{\mbox{\boldmath
$A_I$}}(t_{\rm end})<14.1$ \cite{we}, because ${\mbox{\boldmath
$A_I$}}(t_{\rm end})\neq {\mbox{\boldmath $A_I$}}(0)$ due to
the running.

To find ${\mbox{\boldmath $A_I$}}(t_{\rm end})$ and other couplings
at $t_{\rm end}$ we have to obtain the RG flow interpolating between
the $t=0$ and $t_{\rm end}\simeq 33$. In contrast to the
inflationary stage, the post-inflationary running is very large,
because of its long duration, and requires numerical simulation. In
order to specify more precisely the initial conditions at the top-quark
scale, $t=0$, we take into account the pole mass matching
scheme relating the observable physical masses $M_H$ and $M_t$ to
the running masses $m_H=\sqrt{2\lambda(t)}v$ and (\ref{masses}) (and
relevant couplings) \cite{zucchini,top,espinosa}. One might think
that these subtleties essentially exceed the precision of possible
SM implications in early cosmology. However, the instability bound
on the Higgs mass and its lower bound from the CMB data, which turn
out to be very close to one another, are very sensitive to the pole
mass matching effect, so we include them in numerical simulations.

To have a better comparison with \cite{Wil,BezShap3}, we fix the
$t=0$ initial conditions for the RG equation
(\ref{renorm0})-(\ref{renormZ}) at the top-quark scale $M_t =171$
GeV. For the weak interaction constants $g,g'$ and the strong
interaction constant $g_s$, they read \cite{particle}
    \begin{equation}
    g^2(0) = 0.4202,\  g'^2(0) = 0.1291,
    \ g_s^2(0) = 1.3460,                          \label{initial}
    \end{equation}
where $g^2(0)$ and $g'^2(0)$ are obtained by a simple one-loop RG
flow from the conventional values of $\alpha(M_Z)\equiv
g^2/4\pi=0.0338$, $\alpha'(M_Z)\equiv g'^2/4\pi=0.0102$ at the
$M_Z$-scale, and the value $g_s^2(0)$ at $M_t$ is generated by the
numerical program of \cite{website}\footnote{The analytical algorithm of transition between different scales for $g_s^2$
was presented in  \cite{QCDfromZtotop}.}.
For the Higgs self-interaction constant $\lambda$ and for the Yukawa
top quark interaction constant $y_t$ the initial conditions are
determined by the pole mass matching scheme originally developed in
\cite{zucchini,top} and presented in the following form in the
Appendix of \cite{espinosa}:
    \begin{eqnarray}
    &&\lambda(0) = \frac{M_H^2}{2v^2}(1+2\Delta_H(M_H)),   \label{match}\\
    &&y_t(0) = \frac{\sqrt{2}M_t}{v}(1+\Delta_t(M_H)).       \label{match1}
    \end{eqnarray}
Here, the electroweak vacuum expectation value for the Higgs field is
$v = 246.22$ GeV, and $\Delta_H(M_H)$ and $\Delta(M_H)$ comprise
relevant mass operator corrections in the effective Higgs and top
quark propagators (cf. the relation (\ref{masses}) for masses
without these corrections), which together with $M_H$ depend on
$M_t$, $M_Z$ and the Weinberg angle.

The initial condition $\xi(0)$ is not directly known. It should be
determined from the CMB normalization condition for the amplitude of
the power spectrum (\ref{zeta}) $\Delta_{\zeta}^2\simeq 2.5\times
10^{-9}$ at the pivot point $k_0=0.002$ Mpc$^{-1}$
\cite{WMAPnorm,WMAP} which we choose to correspond to $N\simeq 60$.
As quantum perturbations of the observable $\varphi$ and the new
field (\ref{phi}) are obviously related by
$\zeta_\varphi=\zeta_\phi/Z$, Eq. (\ref{zeta}) immediately yields
the following estimate on the ratio of coupling constants:
    \begin{equation}
    \frac1{Z_{\rm in}^2}\frac{\lambda_{\rm in}}{\xi^2_{\rm in}}
    \simeq 0.5\times 10^{-9}
    \left(\frac{x_{\rm in}\,\exp x_{\rm in}}
    {\exp x_{\rm in}-1}\right)^2              \label{final}
    \end{equation}
at the moment of the first horizon crossing for $N=60$, which we call
the ``beginning'' of inflation and label by $t_{\rm
in}$.\footnote{No modification of the spectral index due to $Z$
occurs in the one-loop RG running, because the difference
$n_s^\varphi-n_s^\phi=2\gamma dt/dN=(\gamma{\mbox{\boldmath
$A_I$}}/48\pi^2)e^x/(e^x-1)$ belongs to the two-loop order.} This
moment, in turn, can be determined from the equations
(\ref{xversusvarphi})--(\ref{x}) and (\ref{quantumscale}) relating
the value of the inflaton field at this moment $\varphi_{\rm in}$ to
the e-folding number $N$ and anomalous scaling ${\mbox{\boldmath
$A_I$}}(t_{\rm in})$. With $\varphi_{\rm in}=M_t\exp(t_{\rm in})$,
this relation takes the form
    \begin{equation}
    t_{\rm in} = \ln \frac{M_P}{M_t}+
    \frac12 \ln\frac{4N}{3\xi_{\rm in}}
    +\frac12
    \ln\frac{\exp x_{\rm in}-1}{x_{\rm in}}.   \label{relation1}
    \end{equation}

The RG equations (\ref{renorm0}) for the six couplings
$(g,g',g_s,y_t,\lambda,\xi)$ with five initial conditions
(\ref{initial})--(\ref{match1}) and the final condition (\ref{final})
at $t_{\rm in}$ defined by Eq.(\ref{relation1}) uniquely determine
the RG flow for given values of the Higgs and top quark masses.
Boundary conditions for this flow are rather involved, because they
go beyond a usual Cauchy problem and even the definition of the
final moment (\ref{relation1}) includes the whole history from $t=0$
to $t=t_{\rm in}$, that is, they are nonlocal in $t$. By iterations,
however, this problem can be numerically solved with {\em
Mathematica}; essential simplifications originate from the fact that
$x_{\rm in}$ turns out to be $O(1)$ (more precisely $\simeq 2$, see
below) and that the last term in (\ref{relation1}) almost always can
be discarded from the expression for $t_{\rm in}$ on top of the
contribution $\ln(M_P/M_t)\simeq 37$.

The RG flow covers also the inflationary stage from the
chronological end of inflation $t_{\rm end}$  to $t_{\rm in}$
(remember that the arrows of the RG parameter $t$ and the physical
time are opposite). As was mentioned above, at the end of inflation
we choose the value $\hat\varepsilon=3/4$
of the slow roll parameter (\ref{varepsilon}),
and $\varphi_{\rm end}=M_P\sqrt{4/3\xi_{\rm
end}}$ under the assumption ${\mbox{\boldmath $A_I$}}(t_{\rm
end})/64\pi^2\ll 1$ which will be justified below. Therefore,
    \begin{equation}
    t_{\rm end} = \ln \frac{M_P}{M_t}
    +\frac12 \ln\frac4{3\xi_{\rm end}}.                  \label{end1}
    \end{equation}
Thus the duration of inflation in units of inflaton field e-foldings
$t_{\rm in}-t_{\rm end}=\ln(\varphi_{\rm in}/\varphi_{\rm end})$ is
very short relative to the post-inflationary evolution $t_{\rm
end}\sim 35$,
    \begin{equation}
    t_{\rm in}-t_{\rm end} = \frac12\ln N
    +\frac12 \ln\frac{\xi_{\rm in}}{\xi_{\rm end}}
    +\frac12
    \ln\frac{\exp x_{\rm in}-1}{x_{\rm in}}\simeq
    \frac12\ln N\sim 2,                  \label{end2}
    \end{equation}
where we took into account that $\xi_{\rm in}\simeq\xi_{\rm end}$
and $x=O(1)$ . This is, of course, typical for large-field
inflationary models with $N\sim\varphi^2/M_P^2$.

This estimate gives the range of our approximation linear in
logarithms. The linearization, say in (\ref{lambda-lin}), implies
the bound $|{\mbox{\boldmath $A_I$}}(t_{\rm end})|\Delta
t/16\pi^2\ll 1$, where $\Delta t<t_{\rm in}-t_{\rm end}\simeq \ln
N/2$, and this approximation holds for
    \begin{equation}
    \frac{|{\mbox{\boldmath $A_I$}}(t_{\rm end})|}{16\pi^2}
    \ll \frac2{\ln N}\simeq 0.5.               \label{linlogappr}
    \end{equation}

\section{Numerical analysis}
The running of ${\mbox{\boldmath $A$}}(t)$ strongly depends on the
behavior of $\lambda(t)$. It is well known that for small Higgs
masses the usual RG flow in SM leads to an instability of the
electroweak vacuum caused by negative values of $\lambda(t)$ in a
certain range of $t$, see e.g. \cite{Sher,espinosa} and references
therein. The same happens with the modified RG for the non-minimally
coupled Higgs field considered here.

\begin{figure}[h]
\centerline{\epsfxsize 12cm \epsfbox{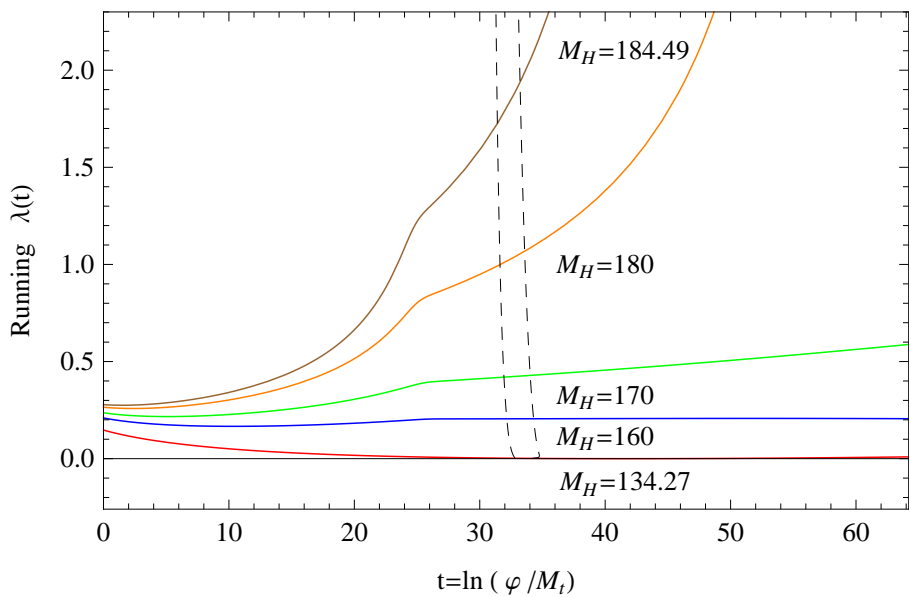}}
\caption{\small Running $\lambda(t)$ for five values of the Higgs
mass above the instability threshold. Dashed curves mark the
boundaries of the inflation domain $t_{\rm end}\leq t\leq t_{\rm
in}$.
 \label{Fig.1}}
\end{figure}

The numerical solution for $\lambda(t)$ in the setting of the
previous section is shown in Fig.1 for five values of the Higgs mass
and the value of top-quark mass $M_t=171$ GeV. The lowest one
corresponds to the critical value
    \begin{equation}
    M_H^{\rm inst}\simeq 134.27\; {\rm GeV}.      \label{criticalmass}
    \end{equation}
This is the boundary of the instability window for which
$\lambda(t)$ bounces back to positive values after vanishing at
$t_{\rm inst}\sim 41.6$ or $\varphi_{\rm inst}\sim 80
M_P$.\footnote{An over-Planckian scale of $\varphi$ does not signify a
breakdown of the semiclassical expansion, because the energy density
$\sim 10^{-10} M_P^4$ stays much below the Planckian value, see
below.} It turns out that the corresponding $\xi(t)$ is nearly
constant and is about $5000$ (see below), so that the factor
(\ref{s}) at $t_{\rm inst}$ is very small, $s\simeq 1/6\xi\sim
0.00005$. Thus the situation is different from the usual Standard
Model with $s=1$, and numerically the critical value turns out to be
higher than the known SM stability bound $\sim 125$ GeV
\cite{espinosa}.

Fig.1 shows that near the instability threshold $M_H=M_H^{\rm inst}$
the running coupling $\lambda(t)$ stays very small for all scales
$t$ relevant to the observable CMB, (\ref{relation1})-(\ref{end1}).
This follows from the fact that the positive running of $\lambda(t)$
caused by the term $(18 s^2+6)\lambda^2$ in $\beta_\lambda$,
(\ref{beta-lambda}), is much slower for $s\ll 1$ than that of the
usual SM driven by the term $24\lambda^2$. For larger $M_H$ this
suppression in the term $18 s^2\lambda^2$ is responsible for the
origin of the bump on the plots at $t\sim 26$ where the value of $s$
drops from 1 to 0 and $\lambda(t)$ continues growing but with a
slower rate.

\begin{figure}[h]
\centerline{\epsfxsize 12cm \epsfbox{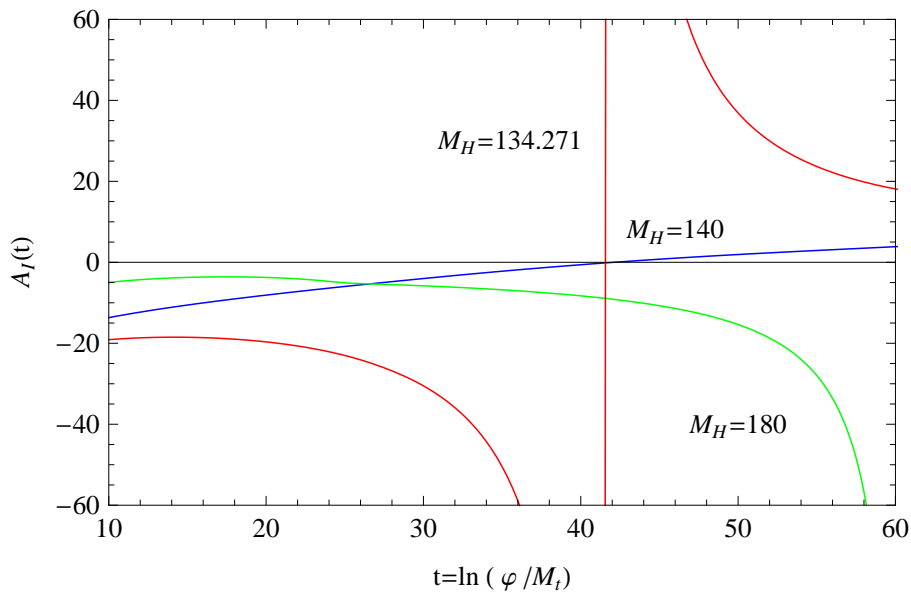}} \caption{\small
Running anomalous scaling for the critical Higgs mass (the red curve
with a vertical segment at the singularity with $t_{\rm inst}\sim
41.6$) and for two masses in the stability domain (blue and green
curves).
 \label{Fig.2}}
\end{figure}
The smallness of $\lambda$ improves, of course, the efficiency of
the perturbation theory. On the other hand, at the critical value of
the Higgs mass we encounter a singularity of both anomalous
scalings ${\mbox{\boldmath $A$}}(t)$ and ${\mbox{\boldmath
$A_I$}}(t)$ at the instability point $t_{\rm inst}$ where
$\lambda(t_{\rm inst})=0$. This is depicted in Fig.~2 for
${\mbox{\boldmath $A_I$}}(t)$ by the red curve corresponding to the
slightly overcritical value $134.271$ GeV (vertical red line and
quasi-hyperbolic curves to the left and to the right of it). Other
overcritical curves with small $M_H>M_H^{\rm inst}$ run through
zero and change the sign of ${\mbox{\boldmath $A_I$}}(t)$ from
negative to positive. For larger $M_H$, when $\lambda$ also becomes
larger, the term $-6\lambda$ in Eq. (\ref{AI}) for ${\mbox{\boldmath
$A_I$}}$ does not let it get positive at any $t$, and
${\mbox{\boldmath $A_I$}}$ starts decreasing after reaching some
maximal negative value (like for the green curve of $M_H=180$ GeV).

\begin{figure}[h]
\centerline{\epsfxsize 10cm \epsfbox{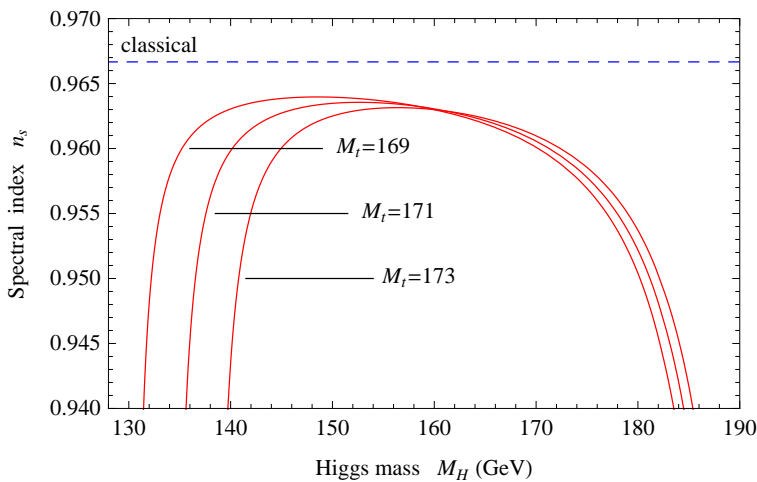}}
\caption{\small The spectral index $n_s$ as a function of the Higgs
mass $M_H$ for three values of the top quark mass.
 \label{Fig.4}}
\end{figure}

An important observation is that for all Higgs masses in the range $M_H^{\rm
inst}=134.27$ GeV $<M_H<185$ GeV the inflation range $t_{\rm
end}<t<t_{\rm in}$ is always below the instability value $t_{\rm
inst}=41.6$ (numerics 
gives that $t_{\rm in}<34.8$), so that from Fig.~2 ${\mbox{\boldmath
$A_I$}}(t)$ is always negative during inflation. Its running
depicted in Fig.~2 explains the main difference from the results of the
one-loop calculations in \cite{we}. ${\mbox{\boldmath $A_I$}}(t)$
runs from big negative values ${\mbox{\boldmath $A_I$}}(0)<-20$ at
the electroweak scale to small but also negative values at the inflation
scale below $t_{\rm inst}$. This makes the CMB data compatible with
the generally accepted Higgs mass range. Indeed, the knowledge of
the RG flow immediately allows one to obtain ${\mbox{\boldmath
$A_I$}}(t_{\rm end})$ and $x_{\rm end}$ and thus find the parameters
of the CMB power spectrum (\ref{ns})--(\ref{running}) as functions of
$M_H$. The parameter of primary interest -- the spectral index -- is
given by Eq. (\ref{ns}) with $x=x_{\rm end}\equiv N{\mbox{\boldmath
$A_I$}}(t_{\rm end})/48\pi^2$ and depicted in Fig.~\ref {Fig.4}. Even
for low values of the Higgs mass above the stability bound, $n_s$ falls
into the range admissible by the CMB constraint existing now at the
$2\sigma$ confidence level (based on the combined WMAP+BAO+SN data
\cite{WMAPnorm}),
    \begin{equation}
    0.94 <n_s(k_0)<0.99.             \label{nsCMB}
    \end{equation}

The spectral index becomes too small (that is, dropping below 0.94)
only for large $x_{\rm end}$ or large negative ${\mbox{\boldmath
$A_I$}}(t_{\rm end})$, which happens only when $M_H$ either
approaches the instability bound or exceeds 180 GeV at the
decreasing branch of the $n_s$ graph. Thus, we get lower and
upper bounds on the Higgs mass, which both follow from a comparison of
the {\em lower} bound of the spectral index in (\ref{nsCMB}) with (\ref{ns}).
Numerical analysis for the corresponding $x_{\rm end}\simeq -1.4$
gives for $M_t=171$ GeV the following range for a CMB-compatible Higgs mass:
    \begin{equation}
    135.62\; {\rm GeV}\lesssim M_H
    \lesssim 184.49\; {\rm GeV}.       \label{CMBmass}
    \end{equation}

Both bounds belong to the nonlinear domain of the equation
(\ref{ns}) because their relevant $x_{\rm end}=-1.4<-1$. However,
their calculation is still in the domain of our linear in logs
approximation, because the quantity $\mbox{\boldmath $A_I$}(t_{\rm
end})/16\pi^2\simeq -0.07$ satisfies the restriction
(\ref{linlogappr}) for $x_{\rm end}=-1.4$. For a smaller
$|\mbox{\boldmath $A_I$}(t_{\rm end})|$ the expression (\ref{ns})
for $n_s$ can be linearized in $x$ and takes a particularly simple
form,
    \begin{equation}
    n_s = 1 -\frac{2}{N} +
    \frac{\mbox{\boldmath $A_I$}(t_{\rm end})}{48\pi^2},\,\,\,\,
    {\mbox{\boldmath $A_I$}}(t_{\rm end})
    \ll \frac{48\pi^2}N\sim 8.                     \label{spectral}
    \end{equation}
It is applicable in a wide range of the Higgs mass in the interior
of the domain (\ref{CMBmass}).

\begin{figure}[h]
\centerline{\epsfxsize 12cm \epsfbox{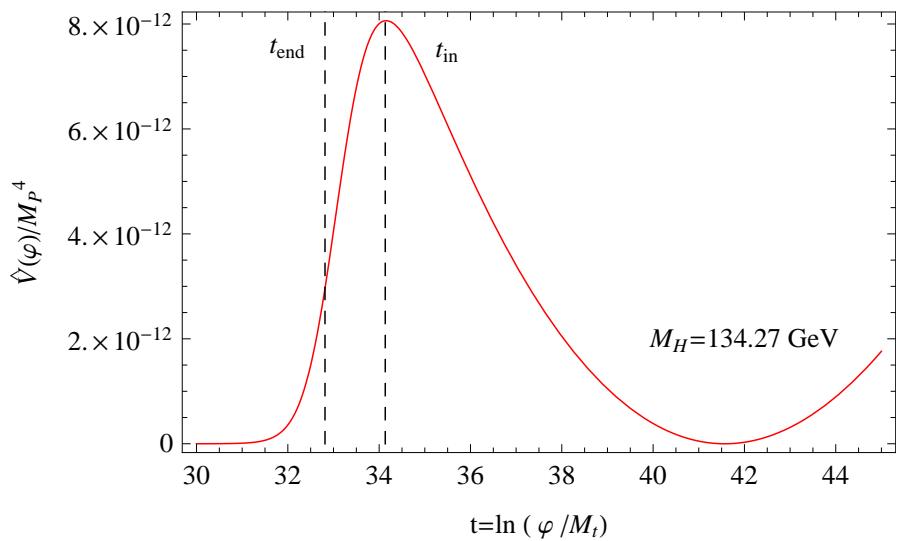}}
\caption{\small The Einstein frame effective potential for the
instability threshold $M_H^{\rm inst}=134.27$ GeV. A false vacuum
occurs at the instability scale $t_{\rm inst}\simeq 41.6$,
$\varphi\sim 80 M_P$, which is much higher than the inflation scale
$\varphi<\varphi_{\rm in}\simeq 0.04 M_P$. A hypothetical inflation domain
(ruled out by the lower $n_s$ CMB bound and the requirement of the
positive slope of $\hat V$ at $t\leq t_{\rm in}$) is marked by
dashed lines. \label{Fig.5}}
\end{figure}

As we see, the upper bound on $n_s$ in (\ref{nsCMB}) does not
generate any restrictions on $M_H$, and it will not effect the $M_H$-range
unless it will be lowered down by future CMB observations to about
0.964 --- the top of the $n_s$-graphs in Fig.~\ref{Fig.4}. The lower
CMB bound in (\ref{CMBmass}) is slightly higher than the instability
bound $M_H^{\rm inst}=134.27$ GeV. This bound depends on
the initial data for weak and strong couplings and, what is even
more important, on the top-quark mass $M_t$ which is known with less
precision. The bound $M_H^{\rm inst}$ given above was obtained for
$M_t=171$ GeV. Below we consider basically this value of the top-quark
mass. Results for the neighboring values $M_t=171\pm2$ GeV are
presented in Fig. \ref{Fig.4} only to show how strongly the plot
gets shifted along the $M_H$ axis. The general pattern of this shift
follows the dependence of $M_H^{\rm inst}$ on $M_t$ -- the
instability bound is larger for a larger top mass, which can be
explained by a negative contribution $-y_t^4\sim -M_t^4$ to
${\mbox{\boldmath $A$}}$, and via ${\mbox{\boldmath $A$}}$ to
$\beta_\lambda$, (\ref{beta-lambda}).

\begin{figure}[h]
\centerline{\epsfxsize 12cm \epsfbox{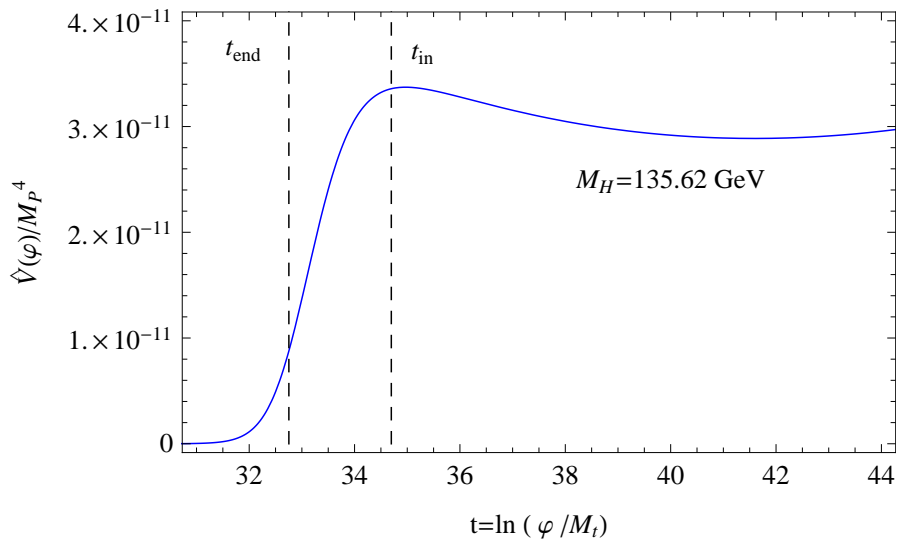}}
\caption{\small Inflaton potential at the lower CMB compatible value
of $M_H$. A metastable vacuum exists at $t\simeq 42$ which is much
higher than the inflation domain at the positive slope of the
potential.
 \label{Fig.6}}
\end{figure}

\begin{figure}[h]
\centerline{\epsfxsize 12cm \epsfbox{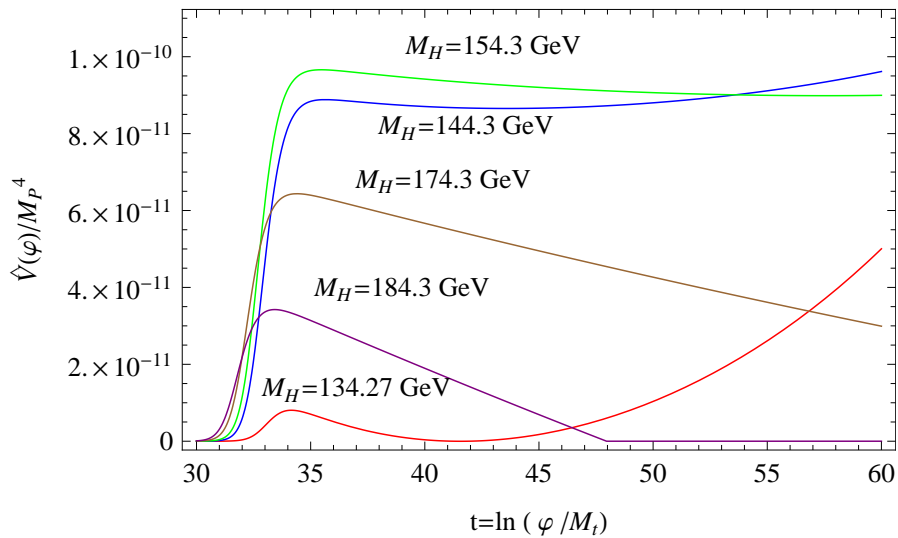}}
\caption{\small The succession of effective potential graphs for
$M_H^{\rm inst}< M_H<184.3$ GeV showing the occurrence of a
metastable vacuum followed for high $M_H$ by the formation of a
negative slope branch. Local peaks of $\hat V$ situated at
$t=34\div35$ grow with $M_H$ for $M_H\lesssim 160$ GeV and start
decreasing for larger $M_H$.
 \label{Fig.7}}
\end{figure}

Viability of the inflation scenario also implies that the system
should safely evolve from the end of inflation to the EW vacuum
through the periods of thermalization (reheating), radiation and
matter domination. The main requirement for this is the possibility
to roll down from $\varphi_{\rm end}=M_t\exp (t_{\rm end})$ to
$\varphi=v$ or the positivity of the effective potential slope
\cite{BezShap1} (we disregard the tunneling scenario). The shape of
the graph of this potential depicted in Fig. \ref{Fig.5} for the
instability threshold confirms the danger of having a negative slope
for $M_H$ close to $M_H^{\rm inst}$ -- the formation of a false
vacuum at the instability scale. At higher Higgs masses till about
160 GeV we first get a family of metastable vacua at scales $\gtrsim
t_{\rm inst}$. An example is the plot for the lower CMB bound
$M_H=135.62$ GeV depicted in Fig. \ref{Fig.6}. For even larger $M_H$
these metastable vacua get replaced by a negative slope of the
potential which interminably decreases to zero at large $t$ (at
least within the perturbation theory range of the model), see
Fig.~\ref{Fig.7}.
However, these metastable vacua and the negative slope of the
potential are not dangerous for the inflationary scenario whose
scales $t_{\rm in}<34.8$ are much lower than the metastability scale
and belong to the positive slope of $\hat V$ to the left of its peak
at $t=34\div35$. Thus, even if the metastable vacuum occurs for low
$M_H$, it exists before the inflation stage probed by current CMB
observations\footnote{The existence of this vacuum can perhaps be
probed by wavelengths longer than that of a pivotal $N\simeq 60$,
but this requires a deeper analysis.}.

\begin{figure}[h]
\centerline{\epsfxsize 12cm \epsfbox{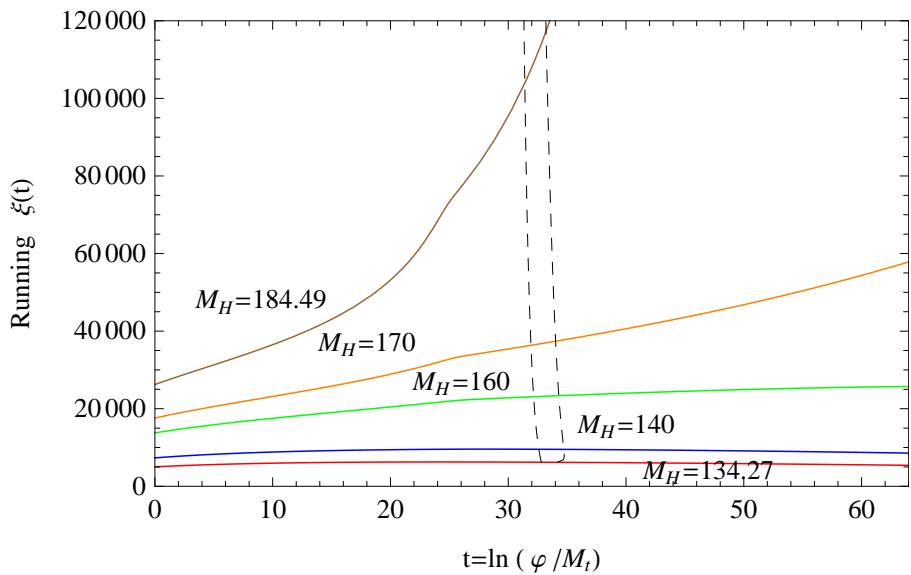}} \caption{\small
Plots of running $\xi(t)$.
 \label{Fig.8}}
\end{figure}
Let us finally focus on the running of $\xi(t)$ depicted for five
values of the Higgs mass in Fig.~\ref{Fig.8}, starting with the lower
bound of the range (\ref{range}) below. It is very slow for low values of
the Higgs mass near the instability threshold, which of course
follows from the smallness of the running $\lambda(t)$ in this domain.
Another property of the $\xi$-behavior is that the normalization of
the power spectrum (\ref{final}) leads to a value $\xi\sim 5000$ for
small Higgs masses, which is smaller than the old estimate $\sim
10^4$ \cite{Spokoiny,SBB,we-scale,BK,efeqmy,BezShap,we}. This is
caused by a decrease of $\lambda(t)$ which at $t_{\rm in}$ becomes
much smaller than $\lambda(0)$ -- an observation made in \cite{Wil}.
This relaxes the situation with the ``unnaturalness'' problem in
this model and brings $\xi$ closer to a very subjective borderline
between ``natural'' and ``unnatural'' values of coupling constants.
For large $M_H$ close to the upper bound of the CMB range, $\xi(t)$
grows like
$\lambda(t)$ to the Landau pole and eventually goes out
of the perturbation theory domain. However, it grows much faster
than $\lambda(t)$, so that $\hat V(\varphi)\sim\lambda(t)/\xi^2(t)$
tends to zero, as shown in Fig. \ref{Fig.7} by the negative slope
branches of the potential.

A final comment of this section concerns the change of numerical
results for a ``two-phase renormalizable'' model of the system when
the $s$-function is replaced by the step function, discussed in
Sect.~4 as a justification for the modified RG scheme. This
replacement leads to a negligible change in numerics -- the
instability threshold remains nearly the same, lower CMB bound on
$M_H$ grows by one in the second digit after the decimal point,
while the upper bound decreases by 1 GeV.

\section{Conclusions and discussion}
We have found that the considered model looks remarkably consistent with CMB
observations in the Higgs-mass range
    \begin{equation}
    135.6\; {\rm GeV} \lesssim M_H
    \lesssim 184.5\;{\rm GeV},                \label{range}
    \end{equation}
which is very close to the widely accepted range dictated by
electroweak vacuum stability and perturbation theory bounds.

Both bounds follow from the lower WMAP bound for the CMB spectral
index $n_s(k_0)>0.94$ (the combined WMAP+BAO+SN data at the pivot
point $k_0=0.002$ Mpc$^{-1}$ corresponding to $N\simeq 60$ e-folds
\cite{WMAPnorm,WMAP}).  The lower bound on $M_H$ is very close to
the instability threshold which in this model equals $M_H^{\rm
inst}\simeq 134.27$ GeV. This is higher than the conventional SM
estimate $\sim 125$ GeV, which is caused by a strong non-minimal
back reaction of gravity and/or the simplicity of our approximation
disregarding higher loop orders.\footnote{It is well known that
despite the perturbative range of coupling constants the two-loop RG
improvement essentially lowers down the EW instability threshold
compared to the one-loop RG running \cite{espinosa}.} Remarkably,
the upper bound in (\ref{range}) does not follow from a demand of a
valid perturbation theory, but rather is imposed by the CMB data.
This CMB mechanism works in a trustable perturbation regime with
$\lambda(t_{\rm in})<2$.

The current upper CMB bound $n_s<0.99$ does not impose restrictions
on the Higgs mass, but it will generate bounds if the observed $n_s$
essentially drops below the ``classical'' value $n_s=1-2/N\simeq
0.967$. Conversely, this model will be falsified if the lower
observational bound on $n_s$ exceeds 0.964 --- the top of the
$n_s$-graphs in Fig.~\ref{Fig.4}.

Our approach represents the RG improvement of our analytical results
in \cite{we}. In fact, here we completely recover the analytic
formalism of \cite{we} for all inflation parameters, which only gets
modified by the RG mapping between the coupling constants at the EW
scale and those at the scale of inflation. A peculiarity of this
formalism is that for large $\xi\gg 1$ the effect of the SM
phenomenology on inflation is universally encoded in one quantity --
the anomalous scaling ${\mbox{\boldmath $A_I$}}$. It was earlier
suggested in \cite{we-scale} for a generic gauge theory (\ref{A}),
and in the SM it is dominated by the contributions of heavy particles --
($W^\pm$, $Z$)-bosons, top quark and Goldstone modes. This quantity
is forced to run in view of RG resummation of leading logarithms,
and this running raises a large negative EW value of
${\mbox{\boldmath $A_I$}}$ to a {\em small negative} value at the
inflation scale. Ultimately this leads to the admissible range of
Higgs masses (\ref{range}) very close to the conventional SM range.

In fact, this mechanism can be regarded as a kind of asymptotic freedom,
for ${\mbox{\boldmath $A_I$}}/64\pi^2$ determines the strength
of quantum corrections in inflationary dynamics \cite{BK,we}.
Usually, asymptotic freedom is associated with the asymptotic
decrease of $\lambda(t)$ to zero. Here, this phenomenon is trickier
because it occurs in the interior of the range (\ref{range}) and
fails near its lower and upper boundaries. For small Higgs masses in
the range (\ref{range}), as it follows from the plots of Fig.
{\ref{Fig.1} and Fig. \ref{Fig.2}, running $\lambda(t)$ gets very
small when approaching the inflationary region $t\sim 32\div 35$ and
generates a large ${\mbox{\boldmath $A_I$}}(t)$ because of its
$1/\lambda$ part induced by vector boson and top quark particles.
For large masses the coupling $\lambda$ grows towards the Landau
pole (though much slower than in nongravitating SM because of the
suppression of the major part of the $\lambda^2$-term of its beta
function). Therefore, ${\mbox{\boldmath $A_I$}}(t)$ becomes big
negative again, this time due to the negative contribution $-6\lambda$
of the Goldstone modes, and falls out of the CMB range at
$M_H\simeq 185$ GeV.

Quantum effects are small only in the middle part of (\ref{range})
with a moderately small $\lambda$ where $n_s$ is close to the
``classical'' limit $1-2/N\simeq 0.967$ for $x\equiv
N{\mbox{\boldmath $A$}}/48\pi^2\ll 1$. Here the original claim of
\cite{BezShap1} on smallness of quantum corrections is right, but
this smallness, wherever it takes place, is achieved via a RG
summation of big leading logarithms.

Qualitatively our main conclusions are close to those of \cite{Wil}
and \cite{BezShap3}, though the RG treatment in these works is very
different from ours. In contrast to them, however, we did not try to
suggest error bars for the Higgs mass range (\ref{range}) and
specify corrections due to the uncertainty in the top quark mass $M_t$.
We only presented all our results for the middle value of the top
mass $M_t=171$ GeV and demonstrated the sensitivity of the
results to this choice by two plots of $n_s$ with two
neighboring values 169 GeV and 173 GeV (see Fig. \ref{Fig.4}). These
error bars as well as the inclusion of the two-loop approximation are,
in our opinion, not reliable. Generally speaking, all the
conclusions of \cite{Wil,BezShap3} and our work still sound too good
to be true in view of the conceptual problems that have not yet been
resolved. These problems were partly mentioned in the criticism of
\cite{Wil} by Bezrukov and Shaposhnikov \cite{BezShap3}, and now we
would like to briefly dwell on them.

As mentioned in \cite{BezShap3}, the RG flow of \cite{Wil}
depends (via the anomalous dimension of the Higgs field $\gamma$) on
the choice of gauge for local $SU(2)\times U(1)$ invariance in the SM.
This inalienable feature of any RG scheme in SM theory was bypassed
in \cite{BezShap3} as follows. The RG flow of all {\rm
gauge-independent} charges of the SM was used at the post-inflationary
stage, while at the inflation stage the effect of running
renormalization of the Higgs field (causing gauge dependence of the
results) was disregarded due to the flatness of the effective potential
$\sim 1/\xi$. In fact, this argument is misleading, because a small
{\em gauge-dependent} slope of the inflation potential
$\sim\dot\varphi$ non-analytically enters the expression for the
e-folding number $N=\int d\varphi H(\varphi)/\dot\varphi$ -- one of
the main ingredients of the inflation parameters $\zeta$ and $n_s$.
Therefore, anyway, this brings an order $O(1)$ gauge dependence into
the expressions for these parameters.

In fact, this is a manifestation of a general problem of the off
shell extension in quantum field theory. Unfortunately, we cannot
formulate the calculation of CMB parameters as on-shell amplitudes
or expectation values of physical observables uniquely defined at
the quantum level. The power spectrum represents the quantum
correlation function calculated at a special field-dependent moment
of time -- the horizon crossing. By using Heisenberg equations of
motion (both for linear quantum perturbations and the nonlinear
background) this function expresses as $\zeta^2=\hat V/24\pi^2
M_P^4\hat\varepsilon$ entirely in terms of the inflaton potential
$\hat V(\varphi)$ and its gradients \cite{MC81,S83,S80}, cf.
Eq.(\ref{varepsilon}) for $\hat\varepsilon$. So the main inflation
observable is the inflaton potential itself, but its quantum version
$\hat V^{\rm eff}$ and its off-shell extension with $(\hat V^{\rm
eff}(\varphi))'\neq 0$ -- the effective potential -- is 
gauge-dependent\footnote{In fact, this is a problem of consistent
transition from $\langle O[V(\varphi)]\rangle$ to $O[V^{\rm
eff}(\langle \varphi\rangle)]$ which brings into game the problem of
gauge dependence in the formalism of the mean field
$\langle\varphi\rangle$ and its effective action.}. Moreover,
off-shell extension of the effective potential (and more generally
effective action) also depends on parametrization of quantum fields.
A particular example of this dependence is given by different results
for CMB characteristics obtained in different conformal frames --
the original Jordan frame versus the Einstein one
\cite{we,BezShap1,BezShap3}. Another important example is the SM
renormalization with different parameterizations of the Higgs
multiplet. In the conventional quantization in Cartesian coordinates
for the complex SU(2) doublet or real O(4) multiplet the theory is
renormalizable and has one set of quantum corrections. In the
spherical coordinate system (used, for example, for the description of
the chiral phase of SM \cite{chiralSM,BezShap3}) the theory is
non-renormalizable and has another set of quantum corrections
\cite{sphericalHiggs}. The S-matrices in both parameterizations
coincide, but their off-shell effective actions are different.

Both off-shell aspects of the above type turn out to be important.
The Einstein frame differs from the Jordan one by a field dependent
conformal rescaling of the metric and a very nonlinear change of the
Higgs field -- radial variable of the Higgs multiplet. Quantization
in these new variables has two major effects which are absent in the original
Jordan frame. One effect is the change of the argument of the
logarithms in quantum corrections, caused by conformal rescaling of
mass parameters \cite{we,BezShap1,ShapRenorm}. This renders these
logarithms nearly constant and not contributing to quantum
corrections. Another effect is the change of coefficients of these
logs \cite{our-ren}. One can show that not only Higgs but also
Goldstone contributions get suppressed in the pre-logarithm
coefficients. Moreover, the Einstein frame used in \cite{BezShap3}
for the description of the chiral phase of the model automatically
leads to the spherical parametrization of the Higgs multiplet, and
this leads to an additional effect. The effective potential of the
$O(N)$ multiplet, calculated in spherical coordinates, does not
contain Goldstone contributions even off shell\footnote{On shell,
$V'=0$, the values of the effective potential of course coincide,
because of a vanishing Goldstone mass (\ref{Goldmass}).}.
Altogether, this leads to a big difference between the results in the Jordan and
the Einstein frame -- there are no large quantum corrections at
high values of $M_H$ when imposing the upper CMB bound on the Higgs range
-- a situation less sensitive to the CMB data, as claimed in
\cite{BezShap3}.

This difference is interpreted in \cite{BezShap3} as an inalienable
theoretical ambiguity of the quantization scheme.
However, we are not ready to completely endorse this statement and
prefer a pragmatic approach which is to try to reduce this
ambiguity as much as possible. Since the cosmological parameters
$\zeta$, $n_s$, etc. are the functionals of the effective
potential, this attempt can be based on the choice of a
preferred parametrization which automatically renders this
potential gauge-independent. Such a parametrization indeed exists
-- this is spherical coordinates for the Higgs multiplet. These
coordinates disentangle the only local gauge invariant observable,
the radial variable
$\varphi=(\varphi^a\varphi^a)^{1/2}=(\Phi^\dagger\Phi)^{1/2}$. The
effective potential for this variable is gauge independent, so
that this solves both parametrization\footnote{Of course, one
might ask the question of effective potential dependence on the
choice of parametrization for Goldstone angles, but we do not even
want to tread into the discussion of this conundrum.} and gauge
independence issues, but of course leaves open the problem of the
conformal frame.

The price we will have to pay for such a resetting of the problem
is that the theory becomes non-renormalizable even in the absence
of gravitational interaction.
Within such a setting it will make sense to include a
two-loop RG improvement\footnote{Two-loop RG improvement and
associated with it error bars in \cite{BezShap3} most likely exceed
available precision for a number of reasons. Not the least of them
is that the two-loop potential was taken from the calculations of
\cite{two-looppotential} in Cartesian coordinates, while its RG
improvement in the chiral phase of inflating SM was based on beta
functions derived in the spherical parametrization of the Higgs
multiplet. Also, the conventional ``Cartesian coordinates'' RG flow
of SM at low energy scales was matched with the chiral RG flow in
spherical parametrization at the inflation stage. The consistency of
this procedure requires checking both at the analytical and
numerical levels.}, establish error bars associated with the current
indeterminacy of the SM and CMB data and the indeterminacy in the
post-inflationary cosmological scenario. All this goes beyond the
goals of the present paper, and will be considered in future
publications. Here we are forced to reconcile with the gauge dependence
of the obtained results. Fortunately, this dependence which enters
through the renormalization function $Z_{\rm in}$ in (\ref{final})
is rather weak. The value of $Z_{\rm in}$ in the Feynman gauge
differs from that of the Landau gauge (used throughout the paper) by
about 6\%, which ultimately effects the value of spectral index in
the third digit after the decimal point.

Finally, in addition to a good match of the spectrum of cosmological
perturbations with the CMB data our model also describes the
mechanism of generating the cosmological {\em background} itself
upon which these perturbations exist. This mechanism is based on the
no-boundary or tunneling cosmological wavefunction
\cite{noboundary,L84,R84,ZS84,tunnel} possibly prescribing initial
conditions for inflation. Within the RG improvement the one-loop
distribution function (\ref{rho}) \cite{norm,we-scale} for the
no-boundary/tunneling state of the Universe is upgraded to
$\rho(\varphi)=\exp\big(\mp\varGamma(\varphi)\big)$, where
$\varGamma(\varphi)$ is the (on-shell) value of the Euclidean
effective action of the quasi-de Sitter instanton, weighting in the
quantum ensemble the member with the initial inflaton value
$\varphi$. The calculation of this action gives
$\varGamma(\varphi)=-24\pi^2 M_P^4/\hat V(\varphi)$ (see \cite{BK})
and the {\em tunneling} distribution function takes the form
    \begin{eqnarray}
    e^{\varGamma(\varphi)}=\exp\left\{
    -24\pi^2\frac{M_P^4}{\hat V(\varphi)}\right\}=
    \exp\left\{-96\pi^2
    \frac{\big(M_P^2+\xi(t)\phi^2\big)^2}
    {\lambda(t)\phi^4}\right\}.               \label{tunneldist}
    \end{eqnarray}

For the CMB range of Higgs masses this distribution features sharp
peaks at the scale $\phi^2_0\simeq\varphi^2_0\simeq-64\pi^2
M_P^2/\xi_0\mbox{\boldmath $A_I$}(t_0)$,\footnote{Naively this
differs by sign from the one-loop scale of inflation
(\ref{quantumscale}), but in fact this is a nontrivial equation for
$\varphi_0$ because the right-hand side of this equality is a
function of $t_0$.} see \cite{QCproject} for details. They exist only
for $\mbox{\boldmath $A_I$}(t_0)<0$ -- which is exactly our case -- and, of
course, correspond to the peaks of the potential in Fig. \ref{Fig.7}
at $t=34\div35$. They have a very small quantum width
$\Delta\varphi/\varphi_0\sim \sqrt\lambda/\xi$ and, therefore, can
be interpreted as generating initial conditions for inflation at the
quantum scale $\varphi_0$ \cite{we-scale,BK,efeqmy}. In the full CMB
range of masses $M_H$ their scale $t_0$ exceeds the $N=60$ CMB
formation scale $t_{\rm in}$, as it should for sake of chronological
succession of initial conditions for inflation and the formation of
CMB spectra (in fact, this is a direct consequence of the positive
slope of $\hat V$ during inflation). Due to the high value of the
exponentiated ratio $\xi^2/\lambda$ in (\ref{tunneldist}) the
probability peak is very distinctive even for the lower CMB value of
$M_H=135.6$ GeV. Moreover, for large $M_H$ close to the upper CMB
bound 185 GeV this peak gets separated from the non-perturbative
domain of large over-Planckian scales due to a fast drop of $\hat
V\sim\lambda/\xi^2$ to zero shown on Fig.~\ref{Fig.7}. This, in turn,
follows from the fact that $\xi(t)$ grows much faster than
$\lambda(t)$ when they both start approaching their Landau pole.
Thus we get a complete {\em tunneling prescription} scenario of the
quantum origin of the inflationary Universe \cite{QCproject}.

To summarize, the inflation scenario driven by the SM Higgs boson
with a strong non-minimal coupling to curvature looks very
promising. This model supports the hypothesis that an
appropriately extended Standard Model can be a consistent quantum
field theory all the way up to the quantum-gravity scale and perhaps explain
the fundamentals of all major phenomena in early and late
cosmology \cite{nuMSM,dark}. Ultimately, it will be the strongly
anticipated discovery of the Higgs particle at LHC and a more
precise determination of the primordial spectral index $n_s$ by
the Planck satellite that might decide the fate of this model.

\acknowledgments{
The authors are grateful to F. Bezrukov, M. Shaposhnikov and O.
Teryaev for fruitful and thought-provoking correspondence and
discussions and also benefitted from discussions with D. Diakonov,
I. Ginzburg, N. Kaloper, I.M. Khalatnikov, D.V. Shirkov, S.
Solodukhin, G.P. Vacca, G. Venturi and R. Woodard. A.B. and A.K.
acknowledge support by the grant 436 RUS 17/3/07 of the German
Science Foundation (DFG) for their visit to the University of
Cologne. The work of A.B. was also supported by the RFBR grant
08-02-00725 and the grant LSS-1615.2008.2. A.K. and A.S. were
partially supported by the RFBR grant 08-02-00923, the grant
LSS-4899.2008.2 and by the Research Programme ``Elementary Particles''
of the Russian Academy of Sciences. The work of C.F.S. was supported
by the Villigst Foundation. A.B. acknowledges the hospitality of LMPT at
the University of Tours. A.S. also acknowledges RESCEU hospitality
as a visiting professor.}

\appendix

\section{Renormalization of potential and non-minimal
coupling terms -- role of Higgs and Goldstone modes}

Here we justify the $s$-factor suppression mechanism due to back
reaction of non-minimally coupled gravitons. For simplicity we
consider only the Higgs multiplet-graviton sector of the model and
the renormalization of the potential and non-minimal coupling terms
of the classical action
    \begin{eqnarray}
    &&S=\int d^4x\,g^{1/2}\left(U(\Phi) R(g)
    -\frac12(\nabla_\mu\Phi^a)^2
    -V(|\Phi|)\right),        \label{model}\\
    &&U\equiv\frac{M_P^2+\xi\Phi^2}2.
    \end{eqnarray}
For generality we work with the $O(N)$ scalar multiplet $\Phi^a$ which
for $N=4$ corresponds to the real representation of the SU(2)
complex Higgs doublet.

For renormalization of non-derivative terms of the action one
expands it in quantum perturbations of the metric and scalar fields
$\sigma^A\equiv(h_{\mu\nu},\sigma)$ on the curved background with a
constant background scalar field $\varphi^a$,
$\Phi^a=\varphi^a+\sigma^a$, $\nabla_\mu\varphi^a=0$. To cancel
nonminimal derivatives in the kinetic term of the action ($\sim
\nabla_\mu\nabla_\nu h^{\mu\nu}$) we pick up the background
covariant DeWitt gauge-breaking term
    \begin{eqnarray}
    &&S_{GB}=-\frac12\,\int d^4x\, g^{1/2} U(\varphi)
    \Big(\nabla^\mu h_{\mu\nu} -\frac12\nabla_\nu h
      -\frac\xi{U(\varphi)}\,\varphi_a\nabla_\nu\sigma^a\Big)^2.
    \end{eqnarray}
The quadratic part of the total action then takes the form
    \begin{equation}
    S_2+S_{GB}=-\frac12\int
    d^4x\,g^{1/2}\,\sigma^A{\mbox{\boldmath$F$}}_{AB}\sigma^B
    \end{equation}
with the following operator acting in the space of gravitons
$h_{\mu\nu}$ and quantum Higgs multiplet $\sigma^a$, the index $A$
running over the range $A=(\mu\nu,a)$,
    \begin{equation}
    {\mbox{\boldmath$F$}}_{AB}=
    \left[\begin{array}{cc}
        \!\!-UG^{\mu\nu,\gamma\delta}(\Box\delta^{\alpha\beta}_{\gamma\delta}
        +P^{\alpha\beta}_{\gamma\delta}
        +\frac{V}{U}\delta^{\alpha\beta}_{\gamma\delta})\!
        &\frac\xi2 g^{\mu\nu}\phi_b\Box+V^{\mu\nu}_b\\
        &\\
        \frac\xi2 \phi_a g^{\alpha\beta}\Box+V^{\alpha\beta}_a
        &\!-\left(\delta_{ab}+\frac{\xi^2\phi^2}
        U n_a n_b\right)\Box+\tilde V_{ab}\!\!
    \end{array}\right],
    \end{equation}
where
    \begin{eqnarray}
    &&P^{\alpha\beta}_{\mu\nu}=2R^{\;\;\;\alpha\;\;\beta}_{(\mu\;\;\nu)}
    +2\delta^{(\alpha}_{(\mu}R^{\beta)}_{\nu)}
    -\delta^{\alpha\beta}_{\mu\nu}R\nonumber\\
    &&\qquad\qquad\qquad
    -g^{\alpha\beta}R_{\mu\nu}-R^{\alpha\beta}g_{\mu\nu}+\frac12
    g^{\alpha\beta}g_{\mu\nu}R,\\
    &&V^{\mu\nu}_a=\frac12 g^{\mu\nu}n_a V'
    +(R^{\mu\nu}-\frac12g^{\mu\nu} R)\,\xi\phi \,n_a,\\
    &&\tilde V_{ab}=m_G^2(\delta_{ab}-n_a n_b)+m_H^2 n_a
    n_b-\xi\delta_{ab}R.
    \end{eqnarray}
Here, the term $-\frac{\xi^2\phi^2}U n_a n_b\Box$ in the kinetic part
of $\sigma^a$ originates from squaring the last term of the gauge,
and $m_G^2=V'/\varphi$ and $m_H^2=V''$ are the Goldstone and Higgs
masses given by Eqs. (\ref{Goldmass}) and (\ref{Higgsmass}) above.
They enter the potential term of the operator as coefficients of the
projectors, respectively, transverse and longitudinal to the direction
of the $O(N)$-multiplet vector of the background scalar field
$n^a=\varphi^a/\varphi$, $\varphi\equiv\sqrt{\varphi^a\varphi^a}$.

The matrix coefficient of $\Box$ in ${\mbox{\boldmath$F$}}_{AB}$
reads
    \begin{equation}
    {\mbox{\boldmath$C$}}_{AB}=
    \left[\begin{array}{cc}
        \,-UG^{\mu\nu,\alpha\beta}\,
        &\,\frac\xi2 g^{\mu\nu}\phi_b\\
        &\\
        \,\frac\xi2 \phi_a g^{\alpha\beta}\,
        &\,-\delta_{ab}-\frac{\xi^2\phi^2}U n_a n_b\,
    \end{array}\right],
    \end{equation}
and its inverse equals
    \begin{equation}
    {\mbox{\boldmath$C$}}^{-1\;AB}=
    \left[\begin{array}{cc}
        -\frac{G_{\mu\nu,\alpha\beta}}U-\frac{1-s}{3U}g_{\mu\nu}g_{\alpha\beta}\,
        &\,\frac{s\xi\phi}U g_{\mu\nu}n_b\\
        &\\
        \,\frac{s\xi\phi}U  n_a g_{\alpha\beta}\,
        &\,-(\delta_{ab}-n_a n_b)-sn_a n_b
    \end{array}\right],
    \end{equation}
so that the operator $\hat
{\mbox{\boldmath$F$}}={\mbox{\boldmath$C$}}^{-1}\!{\mbox{\boldmath$F$}}$
takes the form
    \begin{eqnarray}
    \hat{\mbox{\boldmath$F$}}=\Box\hat{\mbox{\boldmath$1$}}
    +\hat{\mbox{\boldmath$P$}}
    -\frac{\hat{\mbox{\boldmath$1$}}}6R,        \label{operator}
    \end{eqnarray}
where the potential term consists of the two pieces
    \begin{eqnarray}
    \hat{\mbox{\boldmath$P$}}
    =\hat{\mbox{\boldmath$P$}}_\phi
    +\hat{\mbox{\boldmath$P$}}_R.           \label{potterm}
    \end{eqnarray}
The first term is independent of the curvature and all its elements
are $O(1/\xi)$ uniformly for the whole range of $\varphi$ (or for
all scales with $1\geq s>1/6\xi$) except the element
    \begin{equation}
    (\hat{\mbox{\boldmath$P$}}_\phi)^a_b=
    -m_G^2(\delta^a_b-n^a n_b)
    -m_H^2 s\,n^a n_b+n^a n_b\,O\left(\frac1\xi\right).
    \end{equation}
It is dominated by the contribution of the Goldstone mass, and we
explicitly present the term showing the suppression of the Higgs
mass contribution by the factor $s$.

The second term is linear in the curvature and reads
    \begin{equation}
    \hat{\mbox{\boldmath$P$}}_R=\!
    \left[\begin{array}{cc}
        \!P_{\mu\nu}^{\alpha\beta}+\frac16R\delta_{\mu\nu}^{\alpha\beta}
        +O(\frac1\xi)
        &\!\frac{1-s-3s\xi^2\phi}{3U} R g_{\mu\nu}n_b
        -\frac{2\xi\phi}U R_{\mu\nu}n_b\\
        &\\
        -s\xi\phi(R^{\alpha\beta}{-}\frac12 g^{\alpha\beta} R)\,n^a\,
        &\!\!(\hat{\mbox{\boldmath$P$}}_R)^a_b
    \end{array}\right],
    \end{equation}
where
    \begin{equation}
    (\hat{\mbox{\boldmath$P$}}_R)^a_b=
    \left[(\xi+\frac16)(\delta^a_b-n^a n_b)
    +n^a n_b \left(s\xi-\frac16+\frac{s}3\right)\right]R.
    \end{equation}

Now we calculate the one-loop effective action
    \begin{eqnarray}
    S^{\rm 1-loop}=\frac{i}2\,{\rm Tr}\,
    \ln\,\left(\Box\hat{\mbox{\boldmath$1$}}
    +\hat{\mbox{\boldmath$P$}}
    -\frac{\hat{\mbox{\boldmath$1$}}}6R\right)
    \end{eqnarray}
in the approximation linear in spacetime curvature.\footnote{Because
of the constant background scalar field, the Faddeev--Popov ghost
contribution is $\varphi$-independent and does not contribute to the
renormalization of the sector (\ref{model}) of the model.} Its
divergent part is given by the trace of the second Schwinger-DeWitt
coefficient ${\rm tr}\,\hat{\mbox{\boldmath$a$}}_2$ \cite{SchDW}. In
this approximation it is just one half of the square of the full
matrix potential term (\ref{potterm})\footnote{With a non-constant
background $\nabla_\mu\varphi\neq 0$ the operator (\ref{operator})
acquires terms linear in derivatives, which can be absorbed into the
generalized d'Alembertian by the redifinition of the covariant
derivative, $\nabla_\mu\to D_\mu=\nabla_\mu+O(\nabla_\mu\varphi)$.
Then, extra terms in $\hat{\mbox{\boldmath$a$}}_2$ are given by the
square of the curvature associated with the commutator of $[D_\mu,
D_\mu]$ and $D_\mu^2\hat{\mbox{\boldmath$P$}}$. They are responsible
for the renormalization of the kinetic term and curvature squared
terms in the effective action, which we disregard here.}. Since the
$\hat{\mbox{\boldmath$P$}}_\phi$ part of this term commutes with the
rest of the operator (\ref{operator}) in the $1/\xi$-approximation
(in view of the matrix structure and in view of the assumed
constancy of the background $\nabla_\mu\varphi=0$), it can be
considered as a mass matrix
   \begin{eqnarray}
    {\mbox{\boldmath$\hat P$}}_\phi=-{\mbox{\boldmath$\hat M$}}^2,
    \end{eqnarray}
so that
   \begin{eqnarray}
    \hat{\mbox{\boldmath$a$}}_2=\frac12\,\hat{\mbox{\boldmath$P$}}^2
    =\frac12\,\big(\hat{\mbox{\boldmath$M$}}^2\big)^2
    -\hat{\mbox{\boldmath$P$}}_R\hat{\mbox{\boldmath$M$}}^2
    +...\, ,
    \end{eqnarray}
and the one-loop action in dimensional regularization, $d\to 4$,
takes the form explicitly featuring the Coleman--Weinberg structure
of the effective potential,
    \begin{eqnarray}
    &&S^{\rm 1-loop}=\int d^4x\,g^{1/2}\left[\frac1{32\pi^2}
    \left(\frac{1}{2-\frac{d}2}-\mathbb{C}+\ln
    4\pi\right)
    \,{\rm tr}\,\hat{\mbox{\boldmath$a$}}_2\right.\nonumber\\
    &&\qquad\qquad\qquad\quad
    -\frac1{64\pi^2}\,{\rm tr}\,\big(\hat{\mbox{\boldmath$M$}}^2\big)^2
    \Big(\ln\frac{\hat{\mbox{\boldmath$M$}}^2}{\mu^2}-\frac32\,\Big)\nonumber\\
    &&\qquad\qquad\qquad\quad+\left.\frac1{32\pi^2}
    \,{\rm tr}\,\hat{\mbox{\boldmath$P$}}_R
    \hat{\mbox{\boldmath$M$}}^2
    \Big(\ln\frac{\hat{\mbox{\boldmath$M$}}^2}{\mu^2}-1\Big)+...\right].
    \end{eqnarray}
Here, indeed, the second line represents the Coleman-Weinberg
potential and the third line gives the logarithmic renormalization
of the non-minimal coupling, because after calculating the trace it
takes the form $U^{\rm 1-loop}(\varphi)R$. Since
${\mbox{\boldmath$\hat M$}}^2\sim\varphi^2$, the logarithms of the
mass matrix yield, up to subleading corrections, a unit matrix
${\mbox{\boldmath$\hat 1$}}\ln\varphi^2$ and the traces equal
    \begin{eqnarray}
    &&\frac1{64\pi^2}\,{\rm tr}\,\big(\hat{\mbox{\boldmath$M$}}^2\big)^2
    =\frac1{32\pi^2}\Big[\,m_G^4(N-1)+s^2 m_H^4\Big]
    +O\left(\frac1\xi\right)\nonumber\\
    &&\qquad\qquad\qquad\qquad\qquad\qquad\quad
    =\frac{\lambda^2\varphi^4}{128\pi^2}\big(6+18s^2\big)
    +O\left(\frac1\xi\right)                          \label{trace1}\\
    &&\frac1{32\pi^2}{\rm tr}\,\hat{\mbox{\boldmath$P$}}_R
    \hat{\mbox{\boldmath$M$}}^2=
    \frac1{32\pi^2}R\Big[\,m_G^2(N-1)
    + s^2 m_H^2\Big]\,\xi +O(\xi^0)\nonumber\\
    &&\qquad\qquad\qquad\qquad\qquad\qquad\quad
    =\varphi^2 R\,
    \frac{3\lambda}{32\pi^2}(1+s^2)\xi+O(\xi^0).   \label{trace2}
    \end{eqnarray}
These equations show the suppression of the Higgs-mass contribution by
the factors of $s$ and the origin of $6\lambda$ term of the
anomalous scaling in Eq.(\ref{effpot}) (cf. Eq.(\ref{A0})). This
term is entirely due to the contribution of Goldstone modes.
Similarly, they explain the origin of the dominant part of the
constant $\mbox{\boldmath$C$}\sim\xi$ in (\ref{effPlanck}), given by
(\ref{C}).

These equations also underlie the beta functions (\ref{beta-lambda})
and (\ref{beta-xi}) in which we retained only the leading terms in
$\xi\gg 1$, nontrivially depending on $s$-factors. One can check
that subleading terms of (\ref{trace1}) and (\ref{trace2}) above are
uniformly damped by inverse powers of $\xi$ in the full range scales
$1\geq s>1/6\xi$. They originate mainly due to graviton loops
suppressed by powers of effective Planck mass
$1/(M_P^2+\xi\varphi^2)$. The remaining part of the beta functions
(\ref{beta-lambda})--(\ref{beta-xi}) is generated from the heavy
vector boson and quark sectors by adding to the mass matrix
${\mbox{\boldmath$\hat M^2$}}$ their contribution diag
[$\,m_{W_\pm}^2,m_Z^2,m_t^2\,$] and by including the anomalous
dimension terms caused by field renormalization.


\begin{thebibliography}{999}
\bibitem{we-scale}A. O. Barvinsky and A. Yu. Kamenshchik, Phys. Lett. B
{\bf 332} (1994) 270.
\bibitem{BezShap}F. L. Bezrukov and M. Shaposhnikov, Phys. Lett.
B {\bf 659} (2008) 703.
\bibitem{we}A. O. Barvinsky, A. Yu. Kamenshchik and A. A. Starobinsky,
JCAP {\bf 0811} (2008) 021.
\bibitem{BezShap1}
F. L. Bezrukov, A. Magnon and M. Shaposhnikov,
Phys. Lett. B {\bf 675} (2009) 88.
\bibitem{Wil} A. De Simone, M. P. Hertzberg and F. Wilczek,
Phys. Lett. B {\bf 678} (2009) 1.
\bibitem{BezShap3}F. Bezrukov and M. Shaposhnikov, JHEP {\bf 0907} (2009)
089.
\bibitem{Spokoiny} B. L. Spokoiny, Phys. Lett. B {\bf 147} (1984) 39.
\bibitem{FM89} T. Futamase and K.-I. Maeda, Phys. Rev. D {\bf 39}
(1989) 399.
\bibitem{SBB} D. S. Salopek, J. R. Bond and J. M. Bardeen,
Phys. Rev. {\bf D 40} (1989) 1753.
\bibitem{Unruh} R. Fakir and W. G. Unruh, Phys. Rev. D {\bf 41}
(1990) 1783.
\bibitem{BK} A. O. Barvinsky and  A. Yu. Kamenshchik, Nucl. Phys. B
{\bf 532} (1998) 339.
\bibitem{KomatsuFutamase} E. Komatsu and T. Futamase, Phys. Rev. D
{\bf 59} (1999) 0064029.
\bibitem{H82} S. W. Hawking, Phys. Lett. B {\bf 115} (1982) 295.
\bibitem{S82} A. A. Starobinsky, Phys. Lett. B {\bf 117} (1982) 175.
\bibitem{GP82} A. H. Guth and S.-Y. Pi, Phys. Rev. Lett. {\bf 49} (1982)
1110.
\bibitem{noboundary}
J. B. Hartle and S. W. Hawking,
  Phys. Rev.  D {\bf 28} (1983) 2960;
S.W. Hawking, Nucl. Phys. B {\bf 239} (1984)  257.
\bibitem{L84}
A. D. Linde, JETP {\bf 60} (1984) 211; Lett. Nuovo Cim. {\bf 39}
(1984) 401.
\bibitem{R84}
V. A. Rubakov, JETP Lett. {\bf 39} (1984) 107.
\bibitem{ZS84}
Ya. B. Zeldovich and A. A. Starobinsky, Sov. Astron. Lett. {\bf 10}
(1984) 135.
\bibitem{tunnel}
A. Vilenkin, Phys. Rev.  D {\bf 30} (1984) 509.
\bibitem{efeqmy}
A. O. Barvinsky and D. V. Nesterov, Nucl. Phys. B {\bf 608} (2001)
333.
\bibitem{ShapRenorm}M. Shaposhnikov and D. Zenhausern,
Phys. Lett.  B {\bf 671} (2009) 162.
\bibitem{GB08}
J. Garcia-Bellido, D. G. Figueroa and J. Rubio, arXiv: 0812.4642
[hep-ph].
\bibitem{Weinberg}
S. Weinberg, in {\it General Relativity}, ed. S. W. Hawking and W.
Israel (Cambridge University Press, 1979) 790.
\bibitem{norm}
A. O. Barvinsky and A. Y. Kamenshchik, Class. Quant. Grav.  {\bf 7}
(1990) L181.
\bibitem{WMAPnorm} G. Hinshaw et al., Astrophys. J. Suppl. {\bf 180}
(2009) 225.
\bibitem{WMAP}
E. Komatsu et al., Astrophys. J. Suppl. {\bf 180} (2009) 330.
\bibitem{WeinbergQFT}S. Weinberg, {\it The quantum theory of fields. Vol.2.
Modern applications}, CUP, Cambridge, 1996.
\bibitem{particle}
C. Amsler et al., ``Particle Data Group'', Phys. Lett. {\bf B667}
(2008) 1
\bibitem{ColemanWeinberg}S. Coleman and E. Weinberg, Phys. Rev. D
{\bf 7} (1973) 1888.
\bibitem{Woodard} R. P. Woodard, Phys. Rev. Lett. {\bf 101} (2008) 081301,
arXiv:0805.3089 [gr-qc].
\bibitem{MC81}
V. F. Mukhanov and G. V. Chibisov, JETP Lett. {\bf 33} (1981) 532.
\bibitem{S83}
A. A. Starobinsky, Sov. Astron. Lett. {\bf 9} (1983) 302.
\bibitem{S80}
A. A. Starobinsky, Phys. Lett. B {\bf 91} (1980) 99.
\bibitem{our-ren}
A. O. Barvinsky, A. Yu. Kamenshchik and I. P. Karmazin, Phys. Rev. D
{\bf 48} (1993) 3677.
\bibitem{BarbEsp}J. L. F. Barbon and J. R. Espinosa, arXiv:
0903.0355 [hep-ph].
\bibitem{BurgLeeTrott}C. P. Burgess, H. M. Lee and M. Trott, arXiv:
0902.4465 [hep-ph].
\bibitem{BezShap2}F. Bezrukov, D. Gorbunov and M. Shaposhnikov,
JCAP {\bf 0906} (2009) 029.
\bibitem{Kazakov}
D. I. Kazakov, Theor. Math. Phys.  {\bf 75} (1988) 440.
\bibitem{Clarcketal} T.E. Clark, Boyang Liu, S.T. Love, T. ter
Veldhuis, {The Standard Model Higgs Boson-Inflaton and Dark
Matter}, arXiv:0906.5595v1 [hep-ph].
\bibitem{zucchini}
A. Sirlin and R. Zucchini, Nucl. Phys. B {\bf 266} (1986) 389.
\bibitem{top}
R. Tarrach, Nucl. Phys. B {\bf 183} (1981) 384.
\bibitem{espinosa}
J. R. Espinosa, G. F. Giudice and A. Riotto, JCAP {\bf 0805} (2008)
002.
\bibitem{website}
http://www-theory.lbl.gov/\~\ ianh/alpha.html
\bibitem{QCDfromZtotop}D.V. Shirkov, Nucl Phys. {\bf B 371}, 467
(1992); D. V. Shirkov and S. V. Mikhailov, Z. Phys. C {\bf 63}
(1994) 463; R. S. Pasechnik, D. V. Shirkov and O. V. Teryaev, Phys.
Rev.  D {\bf 78} (2008) 071902.
\bibitem{Sher}M. Sher, Phys. Rept. {\bf 179} (1989) 273.
\bibitem{chiralSM}A.C. Longhitano, Phys. Rev. {bf D22} (1980) 1166;
S. Dutta, K. Hagiwara, Q.-S. Yan and K. Yoshida, Nucl. Phys. {\bf
B790} (2008) 111, arXiv:0705.2277.
\bibitem{sphericalHiggs}P.~F.~Kelly, R.~Kobes and G.~Kunstatter,
  Phys.\ Rev.\  D {\bf 50} (1994) 7592
  [arXiv:hep-ph/9406298]; R.~Ferrari,
  arXiv:hep-th/0907.0426.
\bibitem{two-looppotential}C. Ford, I.Jack and D. R. T. Jones, Nucl.
Phys. {\bf B387} (1992) 373; Erratum-ibid {\bf B504} (1997) 551,
hep-ph/0111190.
\bibitem{QCproject}A. O. Barvinsky, A. Yu. Kamenshchik, C. Kiefer,
A. A. Starobinsky, C. Steinwachs, work in progress.
\bibitem{nuMSM}T. Asaka, S. Blanchet and M. Shaposhnikov,
Phys. Lett.  B {\bf 631} (2005) 151; T. Asaka and M. Shaposhnikov,
Phys. Lett.  B {\bf 620} (2005) 17.
\bibitem{dark}M. Shaposhnikov and D. Zenhausern,
  Phys. Lett.  B {\bf 671} (2009) 187.
\bibitem{SchDW}B. S. DeWitt, {\em Dynamical Theory of Groups and Fields},
Gordon and Breach, New York, 1965; A.O. Barvinsky and G.A.
Vilkovisky, Phys. Reports, {\bf 119} (1985) 1.

\end{thebibliography}
\end{document}